\documentclass[pre, preprint, amsfonts]{revtex4-2}

\usepackage{graphicx}
\usepackage{mathtools}
\usepackage{bm}
\usepackage{siunitx}
\usepackage{tabularray}
\UseTblrLibrary{booktabs}
\usepackage{afterpage}
\usepackage[hidelinks]{hyperref}

\begin{document}

\title{Applicability of spatial early warning signals to complex network dynamics}

\author{Neil G. MacLaren$^{1,\dagger}$}
\author{Kazuyuki Aihara$^2$}
\author{Naoki Masuda$^{1,3,4,*}$}
\affiliation{$^1$Department of Mathematics, State University of New York at Buffalo, NY 14260-2900, USA}
\affiliation{$^2$International Research Center for Neurointelligence, The University of Tokyo Institutes for Advanced Study, The University of Tokyo, Japan}
\affiliation{$^3$Institute for Artificial Intelligence and Data Science, State University of New York at Buffalo, USA}
\affiliation{$^4$Center for Computational Social Science, Kobe University, Kobe, 657-8501, Japan}
\affiliation{$^{\dagger}$Present address: U.S. Army Research Institute for the Behavioral and Social Sciences, Fort Belvoir, VA 22060, USA}
\affiliation{$^{*}${\rm naokimas@buffalo.edu}}

\date{\today}

\begin{abstract}
  Early warning signals (EWSs) for complex dynamical systems aim to anticipate tipping points before they occur. While signals computed from time series data, such as temporal variance, are useful for this task, they are costly to obtain in practice because they need many samples over time to calculate. Spatial EWSs use just a single sample per spatial location and aggregate the samples over space rather than time to try to mitigate this limitation. However, although many complex systems in nature and society form diverse networks, the performance of spatial EWSs is mostly unknown for general networks because the vast majority of studies of spatial EWSs have been on regular lattice networks. Therefore, we have carried out a comprehensive investigation of six major spatial EWSs on various networks. We find that the winning EWS depends on tipping scenarios, 
although the coefficient of variation and spatial skewness tend to outperform alternative EWSs. We also find that spatial EWSs behave in a drastically different manner between the square lattice and complex networks and tend to be more reliable for the latter than the former. The present results encourage further studies of spatial EWSs on complex networks.
\end{abstract}

\maketitle

\section{Introduction\label{sec:intro}}

Complex systems display tipping points when there exists some environmental threshold beyond which the system enters a qualitatively different regime \cite{strogatz2024}.
For example, tropical woodland ecosystems may collapse to a relatively barren state as rainfall decreases across a critical threshold \cite{wunderling2022}.
As another example, a novel communicable disease may start to rapidly spread in a population when some environmental conditions are met \cite{pastorsatorras2015}.
Tipping points are in general difficult to anticipate because small changes in driver variables can have markedly different effects on the state of the system \cite{strogatz2024}.
However, a variety of systems display characteristic behaviors in the proximity of a tipping point, and such behaviors have been exploited for developing several early warning signals (EWS) which can anticipate the onset of a tipping point \cite{scheffer2009, chen2012, dakos2012, kefi2014}.

Dynamical systems near a tipping point recover more slowly from a disturbance than those far from a tipping point. This pheonmenon, called critical slowing down, leads to increased autocorrelation and variance in time series data, which are typical EWSs \cite{scheffer2009}. In fact, to calculate these EWSs, many samples from the same element in the system are required in each environmental condition (e.g., a control parameter value in the case of modeling studies) and over several environmental conditions (i.e., some far from a tipping point and others nearer to it) \cite{dakos2012}. 
For example, when samples independently obey an identical normal distribution, emulating one environmental condition, the sample standard deviation, which is a typical EWS, has a standard deviation proportional to $n^{-1/2}$, where $n$ is the number of samples \cite{speed2002, cho2008, vegassanchezferrero2012, masuda2024}. Therefore, ideally, one wants to secure $n=50$ samples or more to reliably estimate the sample standard deviation.
However, in practice, it is often too costly to collect so many samples per environmental condition \cite{biggs2009, dakos2010}, potentially contributing to the lack of consistent EWSs in empirical systems \cite{burthe2016, gsell2016}. Furthermore, if $n$ is large, the environment may drift to a different state in the middle of collecting $n$ samples in the field or experiment. If this is the case, the utility of the EWS computed from the $n$ samples is compromised because the EWS reflects a range of environmental conditions rather than a single one.

Spatial EWSs seek to mitigate this limitation of ``temporal'' EWSs by measuring, in each environmental condition, a single sample from many different elements constituting a complex system, rather than obtaining many samples from one element (or multiple elements) in the system \cite{dakos2010, kefi2014}. We define spatial EWSs as requiring just one sample per element, i.e., $n=1$.
Several proposed spatial EWSs are spatial analogues of temporal EWSs, such as spatial correlation \cite{dakos2010}, spatial variance and skewness \cite{guttal2009}, the power spectrum of a state variable of a spatially extended system \cite{carpenter2010}, and recovery length (i.e., as opposed to recovery time) \cite{dai2013}. Other spatial EWSs have no temporal analogue, such as the distribution of patch sizes in patchy environments \cite{kefi2007}.

EWSs have been probably most vigorously studied for ecological dynamics, many of which take place in physical space. Presumably for this reason, most studies of spatial EWSs have been carried out on spatial regular, grid-like networks modeling two-dimensional ecological landscapes, such as the square lattice with or without periodic boundary conditions \cite{kefi2007, dakos2011, kefi2014, genin2018, sankaran2018, ma2022, genin2024} and
partial differential equations involving space and time \cite{guttal2009, carpenter2010, donangelo2010, dakos2011, buelo2018, majumder2019, pal2022, wang2023, deb2024}. Such a network was also used in a study of EWSs for deforestation transitions \cite{wunderling2022}. However, many other empirical complex systems for which prediction of tipping points is desired have more complex network structure than regular lattices or two-dimensional continuous planes. Examples include epidemic spreading in human and animal populations \cite{pastorsatorras2015}, progression of diseases in general \cite{chen2012} and symptoms of mental disorders in particular \cite{vandeleemput2014, dablander2023}, and
inter-specific population dynamics among animals and plants \cite{bascompte2023}.
Furthermore, even if ecological dynamics occur in a two-dimensional terrain, habitats may be irregularly distributed and heterogeneous \cite{baguette2013, white2018} such that the underlying network is spatial but heterogeneous. 
Therefore, empirical studies of spatial EWSs in ecological systems \cite{kefi2007, litzow2008, dai2013, cline2014, eby2017, litzow2017, butitta2017, rindi2018, majumder2019, tirabassi2023, rindi2024, wang2024} may be better justified if spatial EWSs are shown to be valid on heterogeneous networks rather than on regular lattices. 
However, spatial EWSs have been rarely studied beyond on regular lattices, while notable exceptions exist for complex dynamics models coupling epidemic and opinion dynamics \cite{jentsch2018, phillips2020}.

In sum, despite the need, whether or not and which spatial EWSs perform well on heterogeneous networks is largely unknown. 
In the present study, we comprehensively investigate the performance of six major spatial EWSs on complex networks, compared across dynamics models, environmental parameters, how the tipping point is approached, and networks. We also provide mechanistic understanding of why they work or do not work depending on the situation, and recommended practices based on our numerical results.

\section{Methods\label{sec:methods}}

\subsection{Dynamics\label{sub:dynamics}}

We used the following four stochastic dynamics models on networks: a coupled double-well model, a model of mutualistic species interactions, a susceptible-infectious-susceptible (SIS) model, and a gene-regulatory model. 
For the coupled double-well, mutualistic species, and gene-regulatory dynamics models, we consider two control parameters: the strength of coupling between nodes, $D\geq0$, and a stress parameter, $u$, which can exert negative ($u<0$) or positive ($u>0$) stress uniformly on all nodes. For the SIS model, we only consider $D$, which is more conventionally known as the infection rate, as the sole control parameter because the concept of a uniform stress is not realistic for the SIS model.
The matrix $A = (A_{ij})$ is the adjacency matrix of the network. We assume that the network is connected, undirected, and unweighted.
Each dynamics is simulated with Gaussian noise $\xi_i$ with noise strength $\sigma$. The noise applied to different nodes is assumed to be independent of each other.

A coupled double-well model on networks is given by \cite{kouvaris2012}
\begin{equation} \label{eq:doublewell}
  \text{d}x_i = \left[ -(x_i - r_1)(x_i - r_2)(x_i - r_3) + D \sum_{j = 1}^N A_{ij}x_j + u \right] \text{d}t + \sigma \text{d}\xi_i,
\end{equation}
where $x_i$ is the dynamical state of the $i$th node and represents a numeric attribute, such as the species population size or amount of tree cover; $r_1 < r_2 < r_3$ are parameters which, in the absence of noise and coupling, set the positions of the equilibria;
$N$ is the number of nodes.
The coupled double-well dynamics has been used for modeling various phenomena, including human social movements \cite{brummitt2015}, interacting biological species \cite{lever2020}, and connected climate regions \cite{wunderling2022}.
In the absence of coupling and noise, there are two stable equilibria: a lower equilibrium at $x_i = r_1$ and an upper equilibrium at $x_i = r_3$.
We set $r_1=1$, $r_2=3$, $r_3=5$, $D=0.05$ (if the control parameter is $u$), $u=0$ (if the control parameter is $D$), and $\sigma=0.1$. 
Exceptions to parameter values are provided in the supplementary material.
We initialize this model in either the lower state with $x_i= r_1 = 1$ $\forall i$ or the upper state with $x_i = r_3 = 5$ $\forall i$.
We consider an $i$th node to be in the lower state if $x_i < r_2$ and in the upper state otherwise.

A model of mutualistic species dynamics is given by \cite{gao2016}
\begin{equation} \label{eq:mutualistic}
  \text{d}x_i = \left[B + x_i \left( 1 - \frac{x_i}{K}\right) \left(\frac{x_i}{C} - 1 \right) + D\sum_{j=1}^N A_{ij}\frac{x_ix_j}{\tilde{D} + Ex_i + Hx_j} + u \right]\text{d}t + \sigma \text{d}\xi_i,
\end{equation}
where $x_i$ represents the abundance of the $i$th species, $B$ is a constant incoming migration rate, $K$ is the carrying capacity, $C$ is the Allee constant, and $\tilde{D}$, $E$, and $H$ moderate the effect of the interaction term $x_ix_j$. 
By following \cite{gao2016}, we set $B=0.1$, $K=5$, $C=1$, $\tilde{D}=5$, $E=0.9$, and $H=0.1$.
We use $D=0.05$ (if the control parameter is $u$), $u=-5$ (if the control parameter is $D$), and $\sigma=0.001$. 
In the absence of coupling and dynamical noise, this model has a stable lower state with $x_i = 0$ and a stable upper state with $x_i = K$.
We initialize this model in the upper state with $x_i = 6$ $\forall i$, which is an arbitrary large value representing an extant population \cite{gao2016}. We start the dynamics from the upper state because a common interest in ecology is loss of resilience in the current species assemblage, which is modeled by collapse from the upper state \cite{scheffer2009}.
We consider that any $x_i > C$ and any $x_i \le C$ are in the upper and lower state, respectively. 
To prevent $x_i < 0$ for any node $i$ and time $t$, which is not physical for this model, we set $x_{i} = 0$ whenever our quadrature algorithm produced $x_{i} < 0$ during simulations. We also use the same procedure to prevent $x_i < 0$ for the following two models.

An SIS model on networks in the stochastic differential equation form is given by \cite{pastorsatorras2015}
\begin{equation} \label{eq:SIS}
  \text{d}x_i = \left[ -\mu x_i + D \sum_{j=1}^N A_{ij}(1 - x_i)x_j \right]\text{d}t + \sigma \text{d}\xi_i.
\end{equation}
The node state $x_i$ represents the probability that the $i$th node is infectious
(the $i$th node is susceptible with probability $1-x_i$); $D$ is the infection rate; $\mu$ is the recovery rate.
We use $\mu=1$ and $\sigma = 0.001$. 
In the absence of noise, there is a disease-free equilibrium with $x_i = 0$ $\forall i$, which always exists and is stable when $D$ is below an epidemic threshold value, and an endemic equilibrium in which $x_i > 0$ $\forall i$, which exists and is stable when $D$ is large enough \cite{pastorsatorras2015}. In the presence of noise, some $x_i > 0$ are expected at any value of $D > 0$.
We simulate this model beginning in either the lower (i.e., almost disease-free) state with $x_i = 0.001$ $\forall i$ or the upper (i.e., endemic) state with $x_i = 0.999$.
We consider an $i$th node to be in the lower state if $x_i < 5\sigma$ and in the upper state otherwise.

A model of gene-regulatory dynamics on networks is given by \cite{gao2016}
\begin{equation} \label{eq:genereg}
  \text{d}x_i = \left( -Bx_i^f + D\sum_{j=1}^N A_{ij}\frac{x_j^h}{1 + x_j^h} + u\right)\text{d}t + \sigma \text{d}\xi_i,
\end{equation}
where $x_i$ represents the expression level of the $i$th gene, $B$ and $f$ characterize the behavior of the $i$th gene in isolation, and $h$ controls the interaction of the $i$th and $j$th genes. 
Following \cite{gao2016}, we set $B=1$, $f=1$, and $h=2$. We use $D=1$ (if the control parameter is $u$), $u=0$ (if the control parameter is $D$), and $\sigma=0.001$.
In the absence of noise, this model has an equilibrium at $x_i=0$ $\forall i$, which is stable when $u$ or $D$ is small enough and represents the inactive state, and an active state with $x_i > 0$ $\forall i$, which exists when $u$ or $D$ is sufficiently large. 
We simulate this model from the upper state with $x_i = 2$ $\forall i$ because one is often interested in modeling the loss of resilience of the active state \cite{gao2016}.
We use the same criteria to define the lower and upper states for the gene-regulatory dynamics as we did for the SIS dynamics.

\subsection{Networks\label{sub:networks}}

We used 30 empirical networks and 5 synthetic networks built with different generative models. These networks vary in terms of the number of nodes and edges, the heterogeneity of the degree distribution, and community structure. Separately, we also analyzed a square lattice for comparison purposes. We coerce each network to be undirected and unweighted if it is not, and use only the largest connected component. Details of the individual networks are found in the supplementary material, section~\ref{sec:SInetworks}. %S1

\subsection{Early warning signals\label{sub:ews}}

We compare six types of EWS that require only $n=1$ sample of $x_i$ for each node $i \in \{1, \ldots, N\}$ at any given control parameter value, i.e.,
Moran's $I$, the spatial standard deviation, spatial variance, spatial coefficient of variation (CV), spatial skewness, and spatial kurtosis.
We compute the EWSs using the equilibrium $x_i^*$ values defined below. Furthermore, the information required by these EWSs is modest: all but Moran's $I$ need only the $x_i$ values; Moran's $I$ needs the $x_i$ values and the adjacency matrix of the network. We chose these six EWSs because spatial variance is a straightforward variant of the spatial standard deviation and because, apart from the EWSs specific to ecological populations, the other five EWSs are commonly used in the literature.

Moran's $I$ is defined as follows \cite{legendre1989, okabe2012}:
\begin{equation} \label{eq:moranI}
  I_{\rm M} = \frac{N}{W} \frac{\sum_{i=1}^N \sum_{j=1}^N A_{ij} (x_i - \overline{x}) (x_j - \overline{x})}{\sum_{i=1}^N (x_i - \overline{x})^2},
\end{equation}
where $W \equiv \sum_{i=1}^N \sum_{j=1}^N A_{ij}$, and $\overline{x} = \sum_{i=1}^N x_i / N$.
Moran's $I$ is a measure of spatial correlation because it quantifies the extent to which neighboring sampling sites on the same surface or object (i.e., nodes in the case of networks \cite{okabe2012}) have similar states \cite{legendre1989}. Specifically, $I_{\rm M}$ is the ratio of the normalized cross-products, or covariance, of the node states, $\sum_{i=1}^N \sum_{j=1}^N A_{ij} (x_i - \overline{x}) (x_j - \overline{x})/W$, to their total variance, $\sum_{i=1}^N (x_i - \overline{x})^2/N$. Moran's $I$ is similar to Pearson's correlation coefficient in that it is close to 1 when neighboring $x_i$ have similar values, close to $-1$ when they are dissimilar, and close to 0 when they are not correlated. However, $I_{\rm M}$ can be less than $-1$ or more than $1$ \cite{legendre1989}.

We refer to the sample standard deviation of the node states as the spatial standard deviation, denoted by $s$, i.e., 
\begin{equation} \label{eq:ssd}
  s = \sqrt{\frac{1}{N-1} \sum_{i=1}^N (x_i - \overline{x})^2}.
\end{equation}
Quantity $s$ is the unbiased estimate of the population standard deviation and is often used as an EWS in spatially extended systems
\cite{butitta2017, litzow2017, buelo2018}. We also use the sample variance, $m_2$, where
\begin{equation}
  m_k = \frac{1}{N} \sum_{i=1}^N (x_i - \overline{x})^k
  \label{eq:moment}
\end{equation}
is the $k$th central moment of $x$, and the CV. The CV is the normalized variant of $s$ defined by
\begin{equation}
\text{CV} = \frac{s}{\overline{x}}.
\end{equation}
We note that the CV is often used as a spatial EWS \cite{litzow2008, dai2013, rindi2018, rindi2024}.

Skewness and kurtosis have been proposed as EWS for time series data \cite{dakos2012}, and both have also been used as spatial EWS \cite{guttal2009, butitta2017, buelo2018, wang2023}. 
Skewness and kurtosis are the scaled third and fourth central moments, respectively, of the probability distribution of a random variable $x$. Sample skewness is defined as
\begin{equation}
  g_1 = \frac{m_3}{m_2^{3/2}}
  \label{eq:skew}
\end{equation}
and quantifies the extent to which extreme values tend to appear to the right (positive) or left (negative) side of the mean and approaches 0 as $N\to\infty$ for a symmetric distribution. Skewness may increase (i.e., become more positive) or decrease (i.e., become more negative) as a system approaches a tipping point \cite{kefi2014}. Therefore, to ensure that the desired behavior of $g_1$ as a system approaches a tipping point is encoded into a more positive value, we use $g_1'$ as the EWS, where $g_1' \equiv g_1$ for ascending simulations and $g_1' \equiv -g_1$ for descending simulations (see section \ref{sub:simulations} for the simulation protocol and section \ref{sub:tau} for a similar procedure involving Kendall's $\tau$).

Sample kurtosis is defined as
\begin{equation}
  g_2 = \frac{m_4}{m_2^2}
  \label{eq:kurt}
\end{equation}
and quantifies the magnitude of the extreme values (i.e., how large and how far from the mean) and approaches 3 as $N\to\infty$ for a normal distribution. An adjustment $g_2' = g_2 - 3$ can be used to define excess kurtosis with respect to a standard normal distribution, but we do not use that convention here. Because extreme values should become more common near a tipping point, a larger (i.e., more positive) kurtosis may indicate an approaching tipping point \cite{ma2022}.

\subsection{Simulations\label{sub:simulations}}

\begin{figure}
  \includegraphics[width = \textwidth]{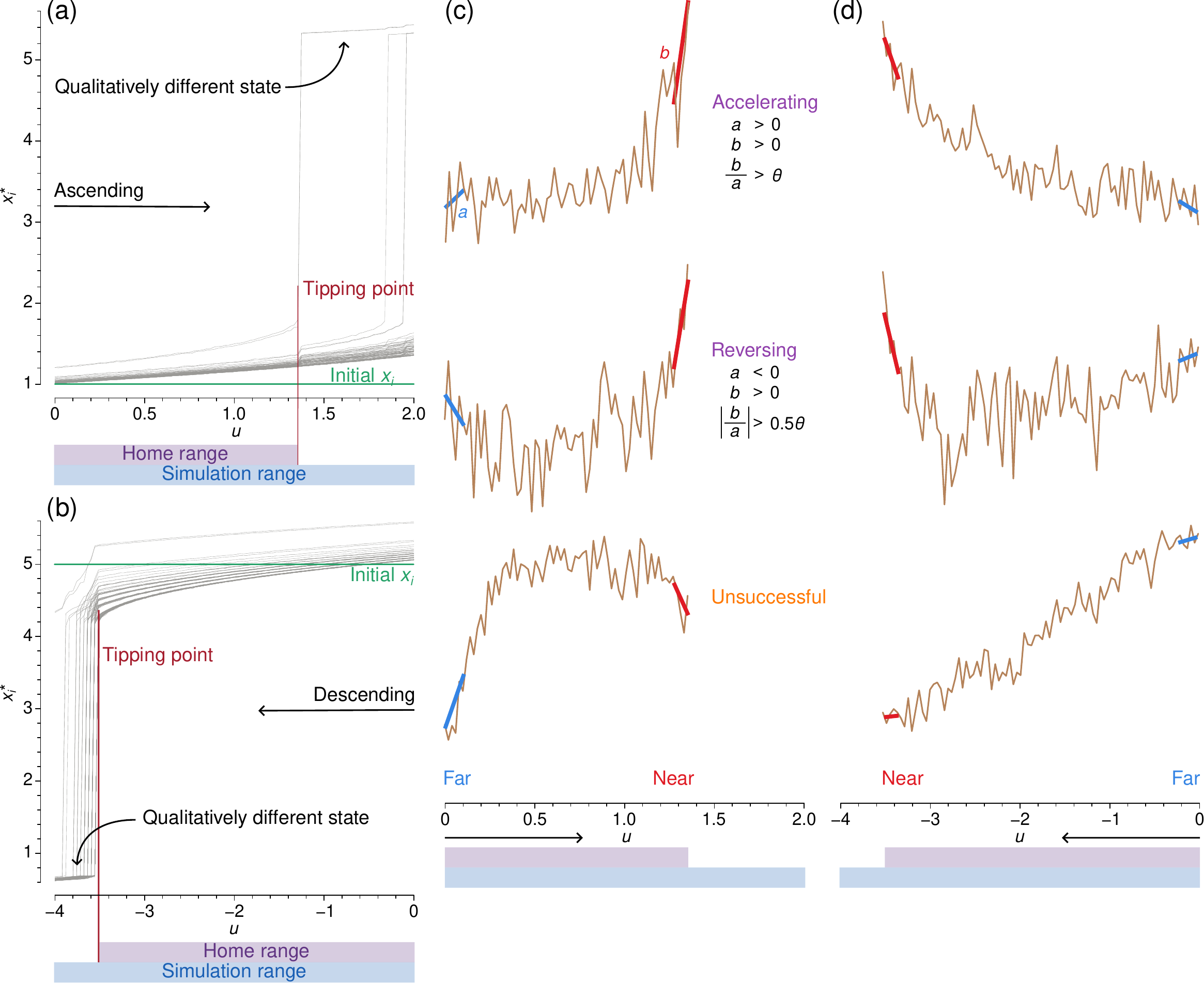}
  \caption{
    Overview of simulations and classification of EWSs. 
    (a) Ascending sequence of simulations of the coupled double-well dynamics on a Barab\'asi-Albert network with 100 nodes. Each gray line shows the $x_i^*$ values at a node. When $u$ is small, the $x_i^*$ are all near their initial value, shown by the green line. As $u$ increases, each $x_i^*$ tends to increase but stays small until the first tipping point, annotated in red. After the first tipping point, at least some nodes move to a qualitatively different state. The home range of the control parameter is defined to be from the first control parameter value, here $u=0$, to the control parameter value immediately before the first tipping point. The simulation range contains the home range and encompasses all sampled values of the control parameter.
    (b) Descending sequence of simulations of the same dynamics on the same network.
    (c) Three hypothetical EWSs (brown lines) for the ascending simulations shown in (a). We classify an EWS by comparing linear regressions of samples of the EWS taken far from (line $a$, blue) and near (line $b$, red) the first tipping point in the home range. We classify the EWS as ``accelerating'', ``reversing'', or ``unsuccessful'' according to the criteria shown.
    (d) Three hypothetical EWS for the descending simulations shown in (b).
  }
  \label{fig:schematic}
\end{figure}

We performed simulations for each combination of dynamics model (i.e., double-well, mutualistic species, SIS, or gene-regulatory), control parameter (i.e., $D$ or $u$), direction (ascending or descending; see below), and network (i.e., one of the 35 networks) and measured the performance of the six EWSs. We refer to a combination of a dynamics model, control parameter, and direction as the simulation condition. As an example, we consider the coupled double-well dynamics with $u$ as the control parameter initialized in the lower state (corresponding to the ``ascending'' simulations as explained below), which altogether specifies a simulation condition, on a Barab\'asi-Albert network (Fig.~\ref{fig:schematic}a). 
We conduct the first simulation with $u=0$, setting $x_{i, t=0}=1$ $\forall i$. We integrate Eq.~\eqref{eq:doublewell} using the Euler-Maruyama method with $\Delta t = 0.01$ for 50 time units.
We consider $x_{i,t=50} = x_i^*$ as an equilibrated value under the dynamical noise, collect one sample of $x_i^*$ from each $i$th node, and calculate the EWSs. Then, we increase $u$ by a small amount and perform the simulation again using the same initial state. The procedure when $D$ is the control parameter is similar.

We iterate these steps to simulate the dynamics and calculate EWSs at $L=100$ evenly spaced control parameter values.
In the present case, we use $100$ values of $u$ in the range $[0, 2]$, which we call the simulation range (Fig.~\ref{fig:schematic}a).
This range of $u$ results in $x_i^* \approx 1$ $\forall i$ when $u$ is smaller than $u \approx 1.35$
and at least some $x_i^*$ in a qualitatively different state, i.e. the upper state (i.e., $x_i^* > 3$), when $u$ is larger.

At large $u$, the statistics of $x_i^*$ of the nodes in the upper state are not useful for predicting which nodes will next make the transition from the lower to the upper state \cite{maclaren2023}.
Therefore, we analyze each EWS only for the control parameter values for which all $x_i^*$ values are near its initial state (i.e., the lower state in the present case). We refer to this restricted parameter range as the home range. 
As shown in Fig.~\ref{fig:schematic}a, the home range is a subset of the simulation range.
Note that the home range is specific to each combination of the simulation condition and network. 

We ensured the following two properties in our simulations. First, the simulation range contains at least one tipping point such that the home range is well-defined (i.e., the last control parameter value in the home range is just before the first tipping point). Second, the simulation range contains sufficiently many control parameter values near and far from the tipping point such that we can assess the performance of the EWSs as described in the following text. To ensure these two properties, we determined the simulation range by trial and error separately for each simulation condition and network. We show the simulation range for each simulation condition and network in the supplementary material (see section~\ref{sec:SIsimulations}). %S2

The sequence of simulations shown in Fig.~\ref{fig:schematic}a begins with each $x_i$ in the lower state and ends when at least some $x_i$ enter the upper state. We call such a sequence of simulations an ascending simulation. By contrast, in Fig.~\ref{fig:schematic}b, we initialize each $x_i$ in the upper state, i.e., $x_{i, t=0}=5$ $\forall i$, and keep reducing $u$ by a small amount for each of the $L$ simulations. We call this type of sequence of simulations a descending simulation.

\subsection{Kendall's $\tau$\label{sub:tau}}

As in prior work, we used the Kendall's rank correlation, denoted by $\tau$, as a performance measure of EWSs \cite{dakos2010, kefi2014}. We computed $\tau$ between the control parameter and the EWS over a range of control parameter values near the tipping event, specifically, the latter half of the home range. We also computed a sign-adjusted Kendall's $\tau'$ value as follows. For dynamics simulated in an ascending sequence, a positive rank correlation indicates that the EWS grows large as the control parameter approaches a bifurcation. In this case, we set $\tau'=\tau$. For dynamics simulated in a descending sequence, a good EWS should grow large as the control parameter becomes smaller (more negative). In this case, we set $\tau'=-\tau$. A larger $\tau'$ value is better.
Both $\tau$ and $\tau'$ range between $-1$ and $1$.

\subsection{Classification of early warning signals\label{sub:classificaion}}

Kendall's $\tau$ is a dominant performance measure for EWSs but with criticisms \cite{boettiger2012, chen2022}. Our numerical simulations produced diverse behavior of the different EWSs as we vary the control parameter, including the case in which the EWS decreases as we approach the tipping point. Given this situation, solely relying on Kendall's $\tau$ would not generate useful comparison between EWSs. Therefore, we developed a classification scheme of EWS as follows.

Consider the example simulation sequence shown in Fig.~\ref{fig:schematic}a, in which we start with the lower state and gradually increase the control parameter, $u$.
Suppose that an EWS responds to the gradual increase in $u$ within the home range of $u$ as shown by the uppermost brown line in Fig.~\ref{fig:schematic}c. This EWS is noisy but remains low when $u < 1$. It then increases progressively rapidly as $u$ approaches the tipping point. This is an ideal case because small values of the EWS reliably indicate that the system is far from transition, whereas large values indicate that the first transition is nearby in terms of $u$. 
We quantify the extent to which an EWS follows this pattern using the trend in the EWS value when it is far from the first transition (slope $a$, blue line in the upper panel of Fig.~\ref{fig:schematic}c) and when it is near (slope $b$, red line). In the upper panel in Fig.~\ref{fig:schematic}c, both $a$ and $b$ are positive and $b$ is substantially larger than $a$. We classify an EWS that behaves in this fashion as successful and say that the EWS is accelerating. We will give the precise definition below.

Alternatively, an EWS may first tend to decrease as $u$ increases when $u$ is far from its tipping point, as in the middle panel of Fig.~\ref{fig:schematic}c. However, if the EWS steadily increases at larger $u$ values near the tipping point, this EWS behavior also reliably indicates that the system is approaching an impending tipping point. Therefore, we also classify this behavior as successful and say that the EWS is reversing.

Many other trends in EWS behavior as a function of the control parameter are possible. For example, the EWS value may initially rise rapidly as $u$ increases far from a tipping point. Then, the EWS may level off or even decrease as $u$ further increases, approaching the tipping point, as in the lower panel of Fig.~\ref{fig:schematic}c. Such an EWS gives a false positive while the system is still far from the bifurcation and a false negative when it is close to the bifurcation. We classify such a behavior, and any other pattern not covered by the accelerating and reversing categories, as unsuccessful. 

We classify EWSs in the same manner when we start from an upper state and gradually decrease the control parameter, as we show in Fig.~\ref{fig:schematic}d.  The lower panel of Fig.~\ref{fig:schematic}d shows a different case of failure (i.e., neither category) from that shown in the lower panel of Fig.~\ref{fig:schematic}c just for demonstration.

To compute $a$, we run an ordinary linear least square regression on the first five values of the control parameter from the home range, in which the independent variable is the control parameter and the dependent variable is the EWS. We compute $b$ in the same fashion but using the last five values of the control parameter in the home range. If $a$ and $b$ are positive and $b/a > \theta$, where $\theta > 1$, we say that the EWS is accelerating. We set $\theta=2$. If $a$ is negative, $b$ is positive, and $|b/a| > 0.5\theta$, then we say that the EWS is reversing. We use $0.5\theta$ to capture trends in an EWS that have increased markedly with respect to an initial negative trend without being too strict. 

\section{Results\label{sec:results}}

We ran numerical simulations on four dynamical system models on networks, i.e., coupled double-well, mutualistic species, SIS, and gene-regulatory models. We performed sequences of simulations of each model across a range of parameter values, forcing a bifurcation, on 35 empirical and model networks. We then computed the six spatial early warning signals for each simulation sequence. To assess the quality of each EWS, we computed $\tau' \in [-1, 1]$, the sign-corrected version of Kendall's rank correlation. We further classified each EWS as accelerating, reversing, or unsuccessful based on the extent to which it showed desirable warning signal behavior far from and near to the tipping point. 
We found that the overall trends of the performance results were similar among the three EWSs based on the spatial variance, i.e., $s$, $m_2$, and $\text{CV}$, and that $\text{CV}$ outperformed $s$ and $m_2$ in most simulations (see the supplementary material, sections~\ref{sec:SItau} % S3
and~\ref{sec:robustness} %S4
for the results). Therefore, we do not show the results for $s$ or $m_2$ in the following text unless we state otherwise.

\subsection{Examples}

\begin{figure}
  \includegraphics[width = 0.7\textwidth]{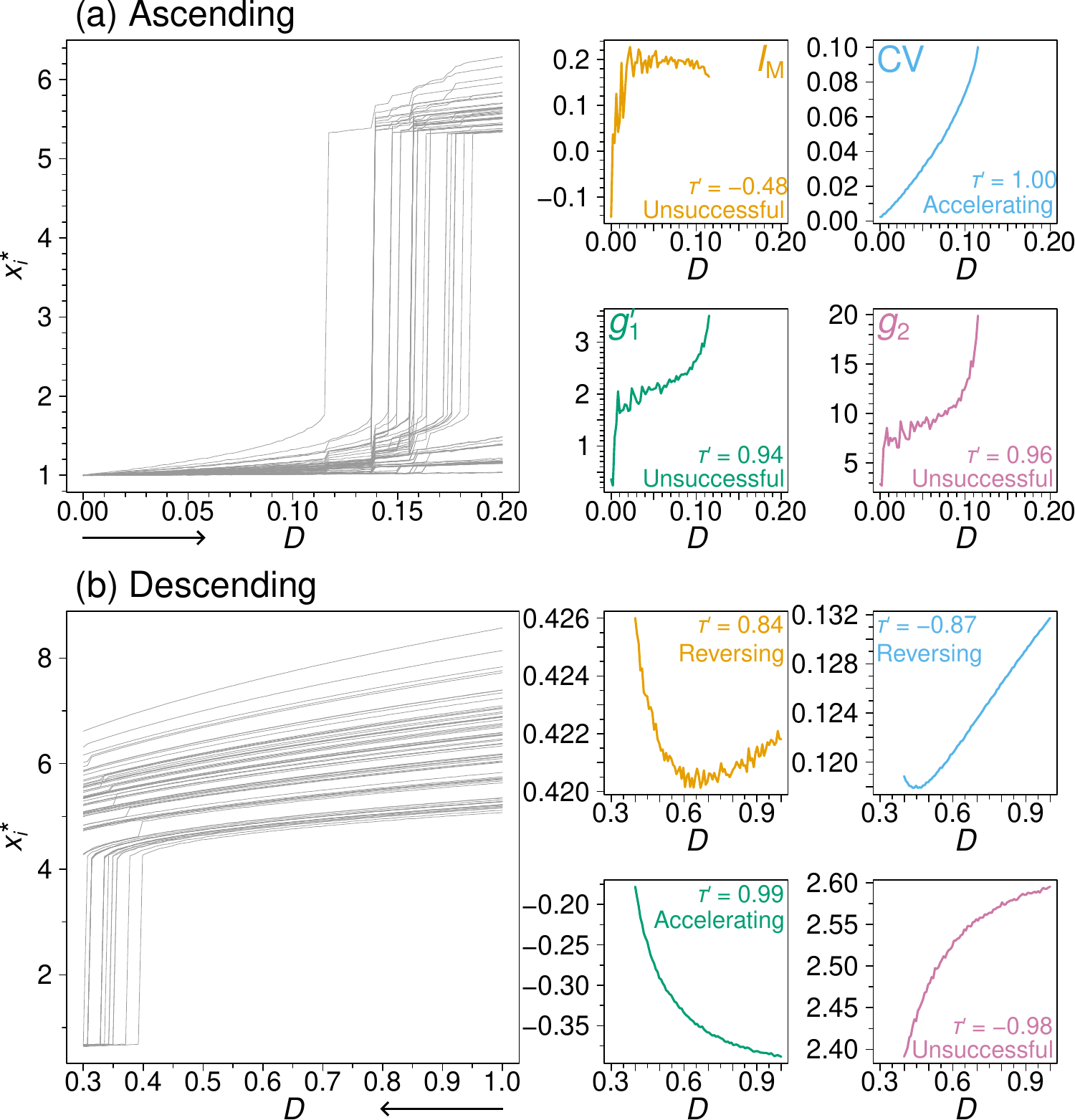}
  \caption{
    Node states and EWSs as a function of $D$ for the coupled double-well dynamics on the drug interaction network.
    (a) Ascending simulations. 
    (b) Descending simulations.
  }
  \label{fig:example-method}
\end{figure}

As an example, let us consider the coupled double-well dynamics (see Eq.~\eqref{eq:doublewell}) on the drug interaction network.
We use the coupling strength, $D$, as the control parameter, which we gradually increase starting from zero (i.e., ascending simulations).
The gray lines in Fig.~\ref{fig:example-method}a represent the $x_i^*$ values for all the nodes as a function of $D$. When $D$ is small, all the nodes are in their lower state (i.e., $x_i^* < 3$). As $D$ gradually increases but is still smaller than the tipping point, each $x_i^*$ becomes larger but still remains in the lower state. The first transition of any node to the upper state occurs around $D \approx 0.115$, and progressively larger values of $D$ result in the transition of more nodes. 

The orange line in Fig.~\ref{fig:example-method}a indicates that $I_{\rm M}$ initially increases rapidly, then levels off and even decreases as $D$ approaches the tipping point. This pattern of $I_{\rm M}$ values is not desirable as EWS and is reflected in the negative sign-adjusted Kendall's rank correlation value ($\tau'=-0.48$); our algorithm classifies $I_{\rm M}$ as ``unsuccessful.''

In contrast, $\text{CV}$, shown by the blue line, grows in an accelerating fashion as $D$ approaches the tipping point, yielding $\tau'=1$ and classification as ``accelerating'', one of our two successful categories. Both $g_1'$ (green) and $g_2$ (magenta) behave similarly to, while more noisily than, $\text{CV}$, except that $g_1'$ and $g_2$ rapidly increase initially as $D$ increases. Although both $g_1'$ and $g_2$ are desirable EWSs in terms of $\tau'$, with $\tau' \approx 1$, our classification scheme classifies them as ``unsuccessful'' due to their initial rapid increase. However, if the simulation starts from a larger value of $D$ while the last value of $D$ remains the same, we can remove the part of the curve in Fig.~\ref{fig:example-method}a where $g_1'$ or $g_2$ rapidly increases. Then, $g_1'$ and $g_2$ would be classified to be successful. See the supplementary material, section~\ref{sec:robustness}, % S4
for results supporting this phenomenon across different networks. Given such a scenario, we advise to refer to both $\tau'$ and our classification scheme in general.

The performance of each EWS depends on the simulation condition (i.e., combination of a dynamics, control parameter, and ascending vs.~descending simulation direction). In Fig.~\ref{fig:example-method}b, we show each $x_i^*$ when we initialize $x_i$ in the upper state and gradually decrease $D$ from $D=1$ (i.e., descending simulations); the dynamics model and network are the same as that used in Fig.~\ref{fig:example-method}a. In this case, $I_{\rm M}$ initially decreases when $D$ decreases and is still far from the tipping point. Then, $I_{\rm M}$ markedly increases as $D$ further decreases and approaches the tipping point. Overall, the trend of $I_{\rm M}$ values is desirable; we obtain $\tau'=0.84$ and classify $I_{\rm M}$ as ``reversing'', one of the two successful categories.

The other EWSs also perform differently from how they did in Fig.~\ref{fig:example-method}a. In Fig.~\ref{fig:example-method}b, $\text{CV}$ performs poorly in terms of $\tau'$; its overall trend is largely linear and monotonically decreasing (rather than increasing) towards the tipping point, yielding $\tau' = -0.87$. We classify $\text{CV}$ as ``reversing'' because $\text{CV}$ increases near the bifurcation. Last, $g_1'$ behaves nearly ideally ($\tau'=0.99$, classified as ``accelerating''), whereas $g_2$ behaves almost conversely ($\tau'=-0.98$, classified as ``unsuccessful'').

In sum, the performance of the different EWSs can substantially vary depending on the direction of the simulations (i.e., ascending versus descending simulations) although the dynamics model and the network are the same. Therefore, we expect that there are various situations in which one EWS may work better than another and vice versa, which we investigate in the following sections.

\subsection{Variability in EWS performance over simulation conditions\label{sub:variability}}

To assess the variation in performance of the different EWSs across conditions, we carried out simulations with each simulation condition on 35 networks.

\begin{figure}
  \includegraphics[width = 0.9\textwidth]{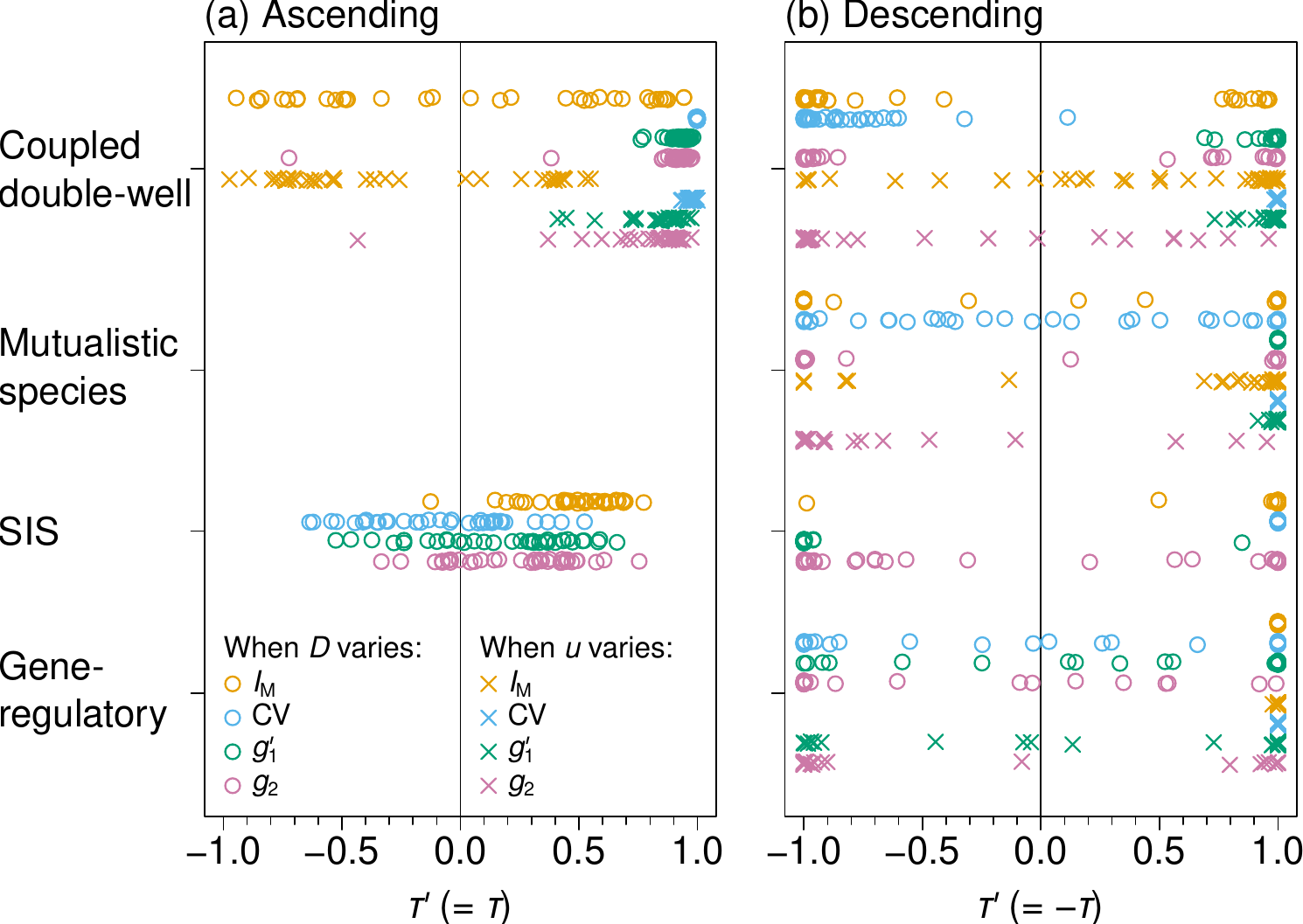}
  \caption{
    Sign-corrected Kendall's $\tau$ values, $\tau'$, for four dynamics models, two control parameters (i.e., $D$, shown by the circles, and $u$, shown by the crosses), and 35 networks. For each set of simulations, $I_{\rm M}$ is shown in orange, $\text{CV}$ in blue, $g_1'$ in green, and $g_2$ in magenta.
    (a) Ascending simulations. (b) Descending simulations.
  }
  \label{fig:taus}
\end{figure}

As an initial analysis, we show the $\tau'$ values for each EWS and simulation condition in Fig.~\ref{fig:taus}. Note that we omitted six simulation conditions being unrealistic (see the Methods section). Therefore, the results for these simulation conditions are missing in Fig.~\ref{fig:taus} (i.e., mutualistic species and gene-regulatory in Fig.~\ref{fig:taus}(a) and variation in $u$ for the SIS). 
We find that $I_{\rm M}$ sometimes performs well (i.e., shown by orange markers near $\tau' = 1$) and sometimes poorly ($\tau'$ much smaller than $1$, including near $-1$). There are many intermediate values of $\tau'$, particularly for ascending simulations (Fig.~\ref{fig:taus}a). On the other hand, $\text{CV}$ performs remarkably well when the control parameter is $u$ (i.e., stress) and under some other simulation conditions in which the control parameter is $D$. Skewness $g_1'$ generally performs well on the coupled double-well and mutualistic species dynamics, and its performance is mixed on the SIS and gene-regulatory dynamics; $g_2$ performs best on ascending simulations of the coupled double-well dynamics and has mixed performance otherwise.

\begin{figure}
  \includegraphics[width = 0.7\textwidth]{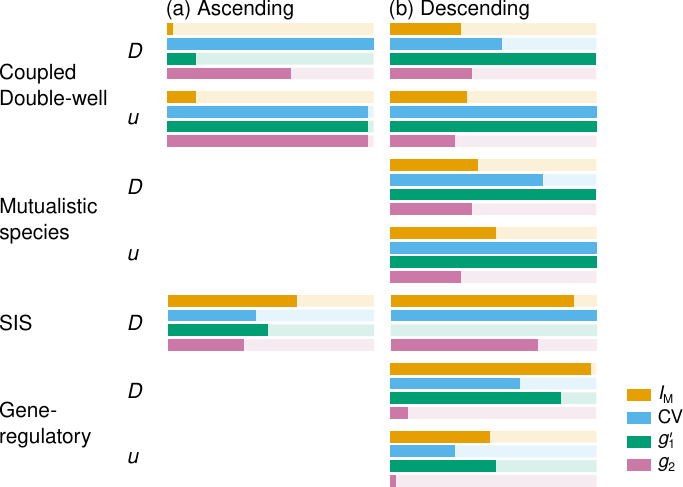}
  \caption{
    Classification of EWSs for four dynamic models on 35 networks. Solid bars indicate the fraction of networks for which the EWS is successful (either accelerating or reversing category). (a) Ascending simulations. (b) Descending simulations.
  }
  \label{fig:performance}
\end{figure}

We applied our classification procedure to further quantify the performance of the different EWSs.
The solid color bars in Fig.~\ref{fig:performance} represent the proportion of networks out of the 35 networks with successful EWSs (i.e., classified as either accelerating or reversing) under each simulation condition. The figure supports the observation made with Fig.~\ref{fig:taus} that the performance of an EWS depends on the simulation condition. For example, $\text{CV}$ is successful for the largest proportion of networks on average among all the EWSs, including $m_2$ and $s$ (see the supplementary material, section~\ref{sec:robustness} %S4
for the results). Instead, $g_1'$ is the best for the coupled double-well and mutualistic species dynamics in descending simulations regardless of the control parameter. Although Fig.~\ref{fig:performance} indicates that $g_1'$ and $g_2$ perform poorly in ascending simulations of the coupled double-well dynamics when the control parameter is $D$, the performance considerably improves if we start simulations from a larger $D$ value (see section~\ref{sec:robustness} %S4
for the results). Moran's $I$, on the other hand, is reliable for the SIS and gene-regulatory dynamics in the descending simulations.

Figure~\ref{fig:performance} suggests that $\text{CV}$ and $g_1'$ perform better than $I_{\rm M}$ and $g_2$ overall. To quantitatively investigate relative performances of the EWSs including $s$ and $m_2$, we assign each EWS a score based on its rank: the best EWS among the six EWSs for a particular simulation condition received 5 points, the second best a 4, and so forth, and worst a 0. In the case of a tie, the tied EWSs receive the average score (e.g., if two EWSs are tied for best, both receive 4.5 points).
Summing this score across the ten simulation conditions confirms that $\text{CV}$ ($32.5$ points) and $g_1'$ ($32$ points) are almost equally the best, followed by $m_2$ ($25$), $I_{\rm M}$ ($23$), $s$ ($22$), and $g_2$ ($15.5$). 
Overall, $\text{CV}$ was categorized as successful in the most cases (76.3\%), followed by $g_1'$ (69.4\%), $m_2$ (54.3\%), $s$ (53.7\%), then $I_{\rm M}$ (48.0\%) and $g_2$ (42.3\%), as shown in the supplementary material, section~\ref{sec:robustness}%S4
. In terms of $\tau'$, it turned out that $g_1'$ was the best, $\text{CV}$ was the second best, and the other four EWSs performed notably worse than $g_1'$ and $\text{CV}$ (see the supplementary material, section~\ref{sec:SItau} %S3
for the results).
Therefore, it appears that $\text{CV}$ and $g_1'$ are the most reliable spatial EWSs among the six EWSs.

In the Introduction, we pointed out that most studies of spatial EWSs were carried out on the square lattice or its continuous variants. 
Therefore, we examined performances of the spatial EWSs on the square lattice with $N=100$ nodes and periodic boundary conditions.
We find that results for the square lattice are substantially different from those for the 35 networks.
Specifically, our algorithm classifies $\text{CV}$ as a successful EWS for the square lattice in $8/10$ simulation conditions, $I_{\rm M}$ in $5/10$ conditions, $g_1'$ in $2/10$ conditions, and $g_2$ in $1/10$ conditions.
The failure of $g_1'$ is in stark contrast with the case of heterogeneous networks. We hypothesized that this last result is because all nodes are structurally equivalent to each other in the square lattice, such that the $x_i$ values are statistically the same for all $i$, yielding the lack of strong asymmetry in the distribution of $x_i$. We have verified that, under the simulation conditions in which $g_1'$ is unsuccessful, $g_1'$ is centered around $0$ as the control parameter varies (see the supplementary material, section~\ref{sec:g1'-square-lattice}%S5
).
Furthermore, except $\text{CV}$, spatial EWSs on the square lattice are typically noisy, leading to smaller $\tau'$ values even when an EWS is classified as successful. 
For $I_{\rm M}$, the average $\tau'$ value is $0.27$ when it is successful on the square lattice and $0.86$ when it is successful on all other networks. For $g_1'$, the average $\tau'$ value is $-0.25$ when it is successful on the square lattice and $0.87$ when it is successful on all other networks. We show examples of the noisiness of EWSs on the square lattice in the supplementary material, section~\ref{sec:SIlattice}%S6
. 
In sum, we conclude that what we know about spatial EWSs on the square lattice does not translate to the case of heterogeneous networks, which further strengthens the need for the present study.

\subsection{Mechanisms of variable performance of early warning signals\label{sub:mechanisms}}

We have shown that different EWSs perform better than others in different simulation conditions.
Focusing on heterogeneous networks, we explore mechanisms behind this observation in this section. Because $g_2$ was the worst performer in all our comparisons, we do not investigate it further.

\subsubsection{$\text{CV}$ \label{subsub:s}}

\begin{figure}
  \includegraphics[width = \textwidth]{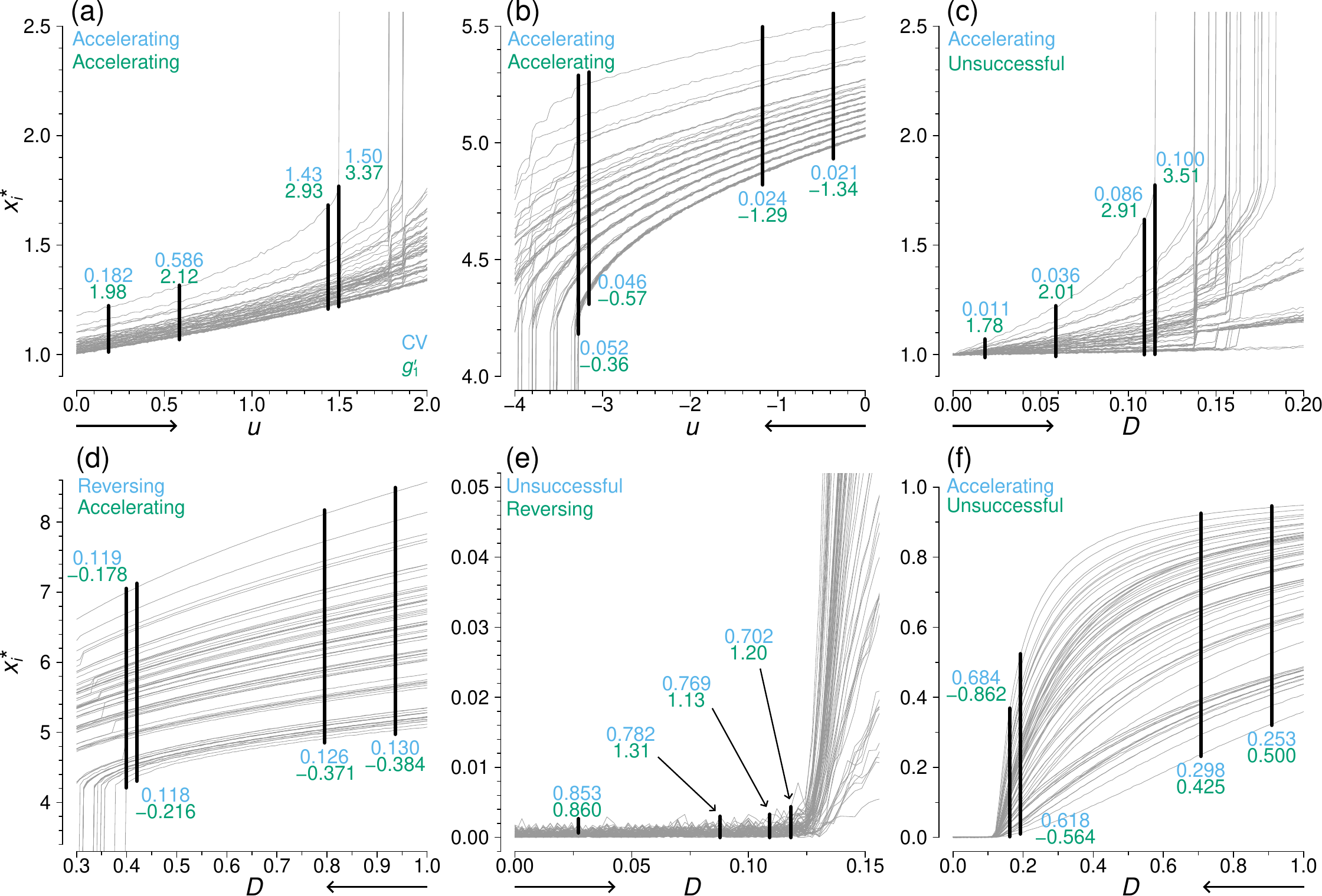}
  \caption{
    Spatial coefficient of variation and skewness as a control parameter varies towards the tipping point.
    Each gray line shows $x_i^*$. The spread of the $x_i^*$ obtained at four control parameter values are shown in black and labeled with the values of $\text{CV}$ (blue) and $g_1'$ (green) at the control parameter values. Classification results are provided in the corresponding colored text. The direction of the simulation sequence is indicated by the arrow beneath the horizontal axis. We used the drug interaction network.
    (a) and (b): Coupled double-well, with $u$ as the control parameter. 
    (c) and (d): Coupled double-well, with $D$ as the control parameter.
    (e) and (f): SIS, with $D$ as the control parameter.
} 
  \label{fig:example-SSD}
\end{figure}

To examine why $\text{CV}$ performs well in some simulation conditions and not in others, let us consider again the coupled double-well dynamics on the drug interaction network used for the demonstration in Fig.~\ref{fig:example-method}. In Fig.~\ref{fig:example-SSD}a, we gradually increase $u$ and observe $x_i^*$ and $\text{CV}$ (and $g_1'$). In this case, $\text{CV}$ increases in an accelerating fashion as $u$ approaches the tipping point, successfully anticipating the saddle-node bifurcation, which the $\text{CV}$ values shown in blue in the figure at four values of $u$ indicate.
Qualitatively the same behavior occurs for all 35 networks.
Figure~\ref{fig:example-SSD}a also indicates that the accelerated increase in $\text{CV}$ is caused by the behavior of the $x_i^*$ values of high-degree nodes (shown by the gray lines with larger $x_i^*$). These nodes receive more input from their neighbors and thus approach a bifurcation earlier than low-degree nodes do \cite{maclaren2023}. These $x_i^*$ values thus separate from those of the smaller-degree nodes near the bifurcation, causing $\text{CV}$ to increase.

In Fig.~\ref{fig:example-SSD}b, we show numerical results under the same simulation condition as in Fig.~\ref{fig:example-SSD}a except that the direction is reversed (i.e., descending simulations). In this case, small-degree nodes separate from the rest as the bifurcation is approached. Nevertheless, their effect on $\text{CV}$ is the same as in the case of Fig.~\ref{fig:example-SSD}a. In other words, the value of $\text{CV}$ grows due to a progressive deviation of the smallest $x_i^*$ values at small-degree nodes as $u$ decreases towards the saddle-node bifurcation.

When the control parameter is $D$, the behavior of $\text{CV}$ is not equivalent in both directions. In the ascending simulations, the $x_i^*$ values of large-degree nodes separate from the remainder (see Fig.~\ref{fig:example-SSD}c; this is the same simulation condition as in Fig.~\ref{fig:example-method}a), which is similar to when the control parameter is $u$ (see Fig.~\ref{fig:example-SSD}a). Therefore, $\text{CV}$ increases in an accelerated manner, successfully anticipating the bifurcation. However, in descending simulations, the $x_i^*$ values of the different nodes come closer together as $D$ decreases towards the bifurcation (see Fig.~\ref{fig:example-SSD}d; also see Fig.~\ref{fig:example-method}b). This result is opposite to the results when the control parameter is $u$ (Fig.~\ref{fig:example-SSD}b).
The reason for this behavior is in that the input from the other nodes to the $i$th node, given by $D\sum_{j=1}^N A_{ij} x_j$, is roughly proportional to the degree of the $i$th node (i.e., the number of $j$ values for which $A_{ij}=1$). Owing to the multiplicative factor $D$, as $D$ becomes smaller, the nodes receive smaller and more homogeneous amounts of input from their neighbors. Therefore, the $x_i^*$ values come closer, reducing $\text{CV}$. Although $\text{CV}$ is still categorized as successful in Fig.~\ref{fig:example-SSD}d, it is because $\text{CV}$ starts to increase very near the bifurcation (i.e., roughly between the two leftmost vertical thick lines in the figure) as $D$ decreases.
The $x_i^*$ values behave similarly as $D$ gradually increases or decreases in the SIS dynamics, which show transcritical bifurcations (see
Fig.~\ref{fig:example-SSD}e and f). However, whether or not $\text{CV}$ is successful in 
Fig.~\ref{fig:example-SSD}e and f does not coincide with the results for the coupled double-well dynamics shown in Fig.~\ref{fig:example-SSD}c and d, respectively. This discrepancy is probably because $\text{CV}$ is a quantity normalized by the mean of $x_i^*$.

\subsubsection{Spatial skewness\label{subsub:g1}}

The causes of the favorable behavior of $g_1'$ as an EWS are related to those of $\text{CV}$ as follows. 

First, $\text{CV}$ and $g_1'$ are successful in Fig.~\ref{fig:example-SSD}a and b for the same reason. EWS $g_1'$ is also successful in  Fig.~\ref{fig:example-SSD}e for the same reason although the increasing trend near the tipping point is weak.
Specifically, as the tipping point is approached, $x_i^*$ of nodes which are closer to the tipping (i.e., high-degree nodes in Fig.~\ref{fig:example-SSD}a and e, and low-degree nodes in Fig.~\ref{fig:example-SSD}b) more rapidly deviate from the remainder, i.e., becoming larger in Fig.~\ref{fig:example-SSD}a and e, and smaller in Fig.~\ref{fig:example-SSD}b. Then, the overall variability of $x_i^*$ grows, captured as an increase in $\text{CV}$ as well as $g_1'$. In this manner, skew increases in ascending simulations at least near the tipping point (i.e., Fig.~\ref{fig:example-SSD}a and e; increasing $g_1'$), driven by the large-degree nodes, whereas skew decreases in descending simulations (i.e., Fig.~\ref{fig:example-SSD}b; this also increases $g_1'$) as the low-degree nodes separate from the other nodes, making the distribution more symmetric. The overall pattern is the same in Fig.~\ref{fig:example-SSD}c except that the initial rapid increase in $g_1'$ causes a misclassification by our algorithm (see Fig.~\ref{fig:example-method}a).

Second, in Fig.~\ref{fig:example-SSD}d, $g_1'$ is clearly successful, whereas $\text{CV}$ is less so. As noted above, there is a tendency for $\text{CV}$ to reverse its decline near the tipping point, while this tendency is weak. In contrast, as the tipping point is approached,
the distribution of $x_i^*$ becomes detectably more symmetric, which successfully increases $g_1'$. This change is primarily caused by a faster decrease in $x_i^*$ at the large-degree nodes rather than by that at the small-degree nodes; the latter was the case in Fig.~\ref{fig:example-SSD}b.

Third, in Fig.~\ref{fig:example-SSD}f, $g_1'$ is unsuccessful because $x_i^*$ is bounded in $[0, 1]$ for the SIS dynamics. Due to this restriction, a faster decrease in $x_i^*$ at the large-degree nodes as the tipping point is approached, which occurs in the case of the coupled-well dynamics to let $g_1'$ perform well (see Fig.~\ref{fig:example-SSD}d), cannot occur in the case of the SIS dynamics.
In this case, the behavior of $g_1'$ is apparently not related to that of $\text{CV}$.

\subsubsection{Moran's $I$}

\begin{figure}
  \includegraphics[width = \textwidth]{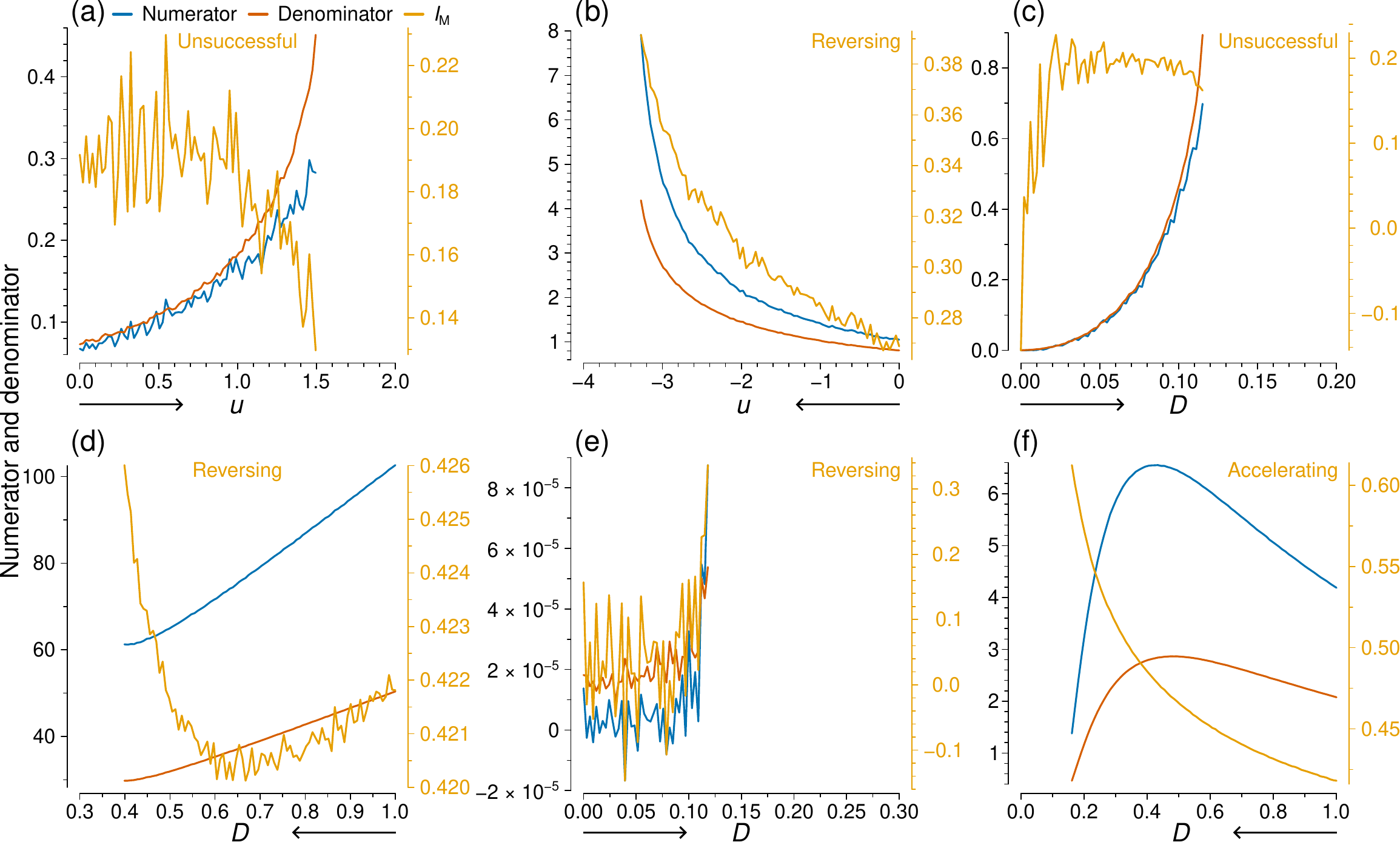}
  \caption{
     $I_{\rm M}$ and the two quantities defining $I_{\rm M}$ for the simulations shown in Fig.~\ref{fig:example-SSD}. In each panel, we show the value of $I_{\rm M}$ (orange line), its numerator (blue line), and its denominator (red line). The direction of the simulation sequence is indicated by the arrow beneath the horizontal axis. The classification result for $I_{\rm M}$ is shown in orange text.
    (a) and (b): Coupled double-well, with $u$ as the control parameter.
    (c) and (d): Coupled double-well, with $D$ as the control parameter.
    (e) and (f): SIS, with $D$ as the control parameter.
  }
  \label{fig:example-I}
\end{figure}

To explore patterns in the behavior of Moran's $I$, we show in Fig.~\ref{fig:example-I} the value of $\sum_{i=1}^N \sum_{j=1}^N A_{ij} (x_i - \overline{x}) (x_j - \overline{x})/W$, which we refer to as the numerator, $\sum_{i=1}^N (x_i - \overline{x})^2/N$, which we refer to as the denominator, and $I_{\rm M}$ for the same six series of simulations as those shown in Fig.~\ref{fig:example-SSD}. Note that $I_{\text{M}}$ is the ratio of the thus defined numerator to the denominator.
The numerator is a covariance function. The denominator is a variance function and the same as $m_2$.

In Fig.~\ref{fig:example-I}a and c,
both the numerator and the denominator of $I_{\rm M}$ increase in an accelerating manner, which is desirable as an EWS.
However, $I_{\rm M}$, which is the ratio of the two quantities, at first remains around the same average value and then declines as $u$ further increases. Therefore, the ratio of two apparently successful EWSs generates a poor EWS. In fact, $I_{\rm M}$ is classified as unsuccessful in these two cases. In contrast, in Fig.~\ref{fig:example-I}b and e, the ratio of two increasing, accelerating quantities, which would individually make desirable EWSs, leads to a successful EWS.
Conversely, in Fig.~\ref{fig:example-I}d and f, $I_{\rm M}$ is a successful EWS, but neither of its components is.
We conclude that the success or failure of $I_{\rm M}$ depends on the intricate balance between the numerator and denominator and that $I_{\rm M}$'s performance is not much linked to the performance of the numerator or denominator. 

\section{Discussion\label{sec:discussion}}

We comprehensively analyzed the performance of major spatial EWSs across dynamics models, control parameters, simulation directions, and networks.
We also gave mechanistic insights into the reasons why good performers work well under some simulation conditions but not in others
(see Figs.~\ref{fig:example-SSD} and \ref{fig:example-I}).
Figure~\ref{fig:performance}, which is our main result, indicates that there is no clear overall winner. However, the figure shows some tendency, which we propose as a recommended practice. When the networked complex system shows saddle-node bifurcations (i.e., coupled double-well and mutualistic species dynamics), $g_1'$ performs the best overall. Furthermore, if the system is known to transit from the lower state to the upper state, $\text{CV}$ outperforms $g_1'$. However, in ecological and deforestation dynamics, sudden large drops from an upper to lower state, which may reflect a saddle-node bifurcation, are of practical interest, corresponding to mass extinction and deforestation, respectively. Therefore, we recommend $g_1'$ for these applications. In contrast, when the observables do not show sudden large jumps at tipping points (i.e., SIS and gene-regulatory dynamics), $I_{\rm M}$ is the best at large. However, all the spatial EWSs including $I_{\rm M}$ perform relatively poorly in two of the four simulation conditions without sudden large jumps (i.e., ascending simulations in the SIS dynamics and descending simulations with $u$ as the control parameter in the gene-regulatory dynamics). Therefore, we conclude that tipping points of these dynamics are relatively difficult to anticipate. We emphasize that $I_{\rm M}$, which is a commonly used spatial EWS, performs well only under specific simulation conditions (i.e., combination of dynamics with no sudden large jump, $D$ as the control parameter, descending simulations). Furthermore, computation of $I_{\rm M}$ needs the adjacency matrix of the network, which is not the case for $\text{CV}$, $g_1'$, or $g_2$.

For the square lattice, some studies report that both spatial correlation and spatial variance anticipate tipping points acceptably well \cite{kefi2014, butitta2017, eby2017, sankaran2018, majumder2019}, whereas others report that spatial correlation outperforms spatial variance \cite{buelo2018, xu2015, wang2024} or vice versa \cite{cline2014, pal2022}. Still others report that alternatives such as spatial skewness and spatial kurtosis \cite{ma2022, wang2023}, recovery length \cite{dai2013, rindi2018, rindi2024}, information measures \cite{phillips2020, deb2024, tirabassi2024}, or a combination of measures \cite{guttal2009} are more reliable. Other reports say that different spatial EWSs have better or worse performance even given the same dynamics model, depending on, for example, the bifurcation type \cite{dakos2011}, direction from which the bifurcation is approached \cite{buelo2018}, or parameter values \cite{dakos2010, genin2018, genin2024}. We showed that spatial EWSs show a diversity of results on heterogeneous networks as well. However, we also showed that one of the two best performers on heterogeneous networks (i.e., $g_1'$) performs poorly on the square lattice. Therefore, the findings available for spatial EWSs on the square lattice do not much help us understand how they behave in heterogeneous networks. This paper is a first comprehensive report for heterogeneous networks. Follow-up studies using other types of spatial EWSs and a larger variety of dynamics models and heterogeneous networks, including examining the effect of network structure, are warranted for future work. Other types of spatial EWSs include
spectral reddening \cite{carpenter2010},
mutual information \cite{phillips2020}, spatial permutation entropy \cite{tirabassi2024}, and
patch-size distribution and patch shape \cite{kefi2007, kefi2014}.

A side contribution of this paper is a new classification method for EWSs into two successful categories and one unsuccessful category. Our proposal was motivated by the insufficiency of the predominantly used performance measure for EWSs, i.e., Kendall's $\tau$, which was pointed out in the literature \cite{boettiger2012, chen2022}.
Our measure aims to capture how the EWS nonlinearly changes towards a tipping point and regards that an accelerated increase nearer to the tipping point implies a better signal. However, the behavior of EWSs is probably more diverse than what our classification or Kendall's $\tau$ can capture. We did not provide stopping criteria either, because it is not a focus of the present work. Further work is needed for better assessments of the quality of EWSs, including temporal EWSs.

Our numerical results suggested that well-performing spatial EWSs may depend on the type of tipping points. This possibility is worth further investigation.
In real applications, one does not typically know what type of tipping point is being approached. In our opinion, however, a small number of bifurcation types, in particular, saddle-node, transcritical, and Hopf bifurcations, seem to be able to represent a large number of empirical systems. Therefore, we propose to focus on investigating these major types of bifurcations to further understand spatial EWSs on networks.
Use of various other dynamical systems on networks and the topological normal form of dynamical systems may be fruitful to this end. Furthermore, recently developed theory that clarified universal features of the interplay between network structure and dynamics on networks (e.g., \cite{barzel2013, hens2019, barzon2024}) may be useful for theoretical investigation of spatial EWSs on networks. Also important is to clarify the behavior of spatial EWSs in various false positive and false negative scenarios (e.g., when there is a regime shift, or discontinuous jumps in $x_i$, but critical slowing down is absent) \cite{boettiger2012, boettiger2013, bury2021, obrien2023, tirabassi2024}. We have verified that the present spatial EWSs do not anticipate sudden and large jumps of $x_i$ in two major scenarios (described in \cite{boettiger2012, boettiger2013}) in which the jumps do not involve critical slowing down. See the supplementary material, section~\ref{sec:no-csd}%S7
, for the results. On the other hand, the spatial EWSs can often anticipate continuous bifurcations such as in the SIS dynamics, as expected. Because epidemic outbreaks are certainly a desired application of EWSs \cite{southall2021}, the size or continuity of the jump should not be the only main criterion to characterize regime shifts. Furthermore, if a sudden change in the value of a system parameter induces a sudden regime shift (see section~\ref{sec:no-csd}%S7
), by construction, there should be no useful EWS before the parameter value changes. Therefore, we need further discussion on which type of system changes should be successfully anticipated, advancing the existing literature on false positives and negatives of EWSs \cite{boettiger2012, boettiger2013}.

Although spatial EWSs were originally proposed for spatial regular networks (i.e., the square lattice and its space-continuous limit), the present results suggest that these EWSs are rather more promising for heterogeneous networks, which most complex empirical networks are. Therefore, this work motivates further work on spatial EWSs for complex networks and their applications to empirical data.

\section*{Data accessibility}

Data and code to reproduce these analyses can be found at \url{https://zenodo.org/records/14681608}.%\url{https://github.com/ngmaclaren/spatialEWS}.

\section*{Author contributions}

N.G.M.: investigation, methodology, software, writing--original draft, writing--review \& editing; K.A.: funding acquisition, investigation, writing--review \& editing; Na.M.: conceptualization, funding acquisition, investigation, methodology, project administration, supervision, writing--original draft, writing--review \& editing.

\section*{Acknowledgments}

This work was performed in part at the Center for Computational Research, State University of New York at Buffalo.

\section*{Funding}

K.A. and Na.M. acknowledge support from the Japan Science and Technology Agency (JST) Moonshot R\&D (grant no. JPMJMS2021).
K.A. also acknowledges support from
the Institute of AI and Beyond of The University of Tokyo,
the International Research Center for Neurointelligence (WPI-IRCN) at The University of Tokyo Institutes for Advanced Study (UTIAS),
JSPS KAKENHI (grant no. JP20H05921),
and the Cross-ministerial Strategic Innovation Promotion Program (SIP), the third period of SIP (grant nos. JPJ012207 and JPJ012425).
Na.M. also acknowledges support from the National Science Foundation (grant no. 2052720),
the National Institute of General Medical Sciences (grant no. 1R01GM148973-01),
and JSPS KAKENHI (grant nos. JP 23H03414, 24K14840, and 24K030130).

\bibliographystyle{unsrt}
\bibliography{refs.bib}

\begin{thebibliography}{100}

\bibitem{strogatz2024}
S.~H. Strogatz.
\newblock {\em Nonlinear dynamics and chaos: With applications to physics,
  biology, chemistry, and engineering}.
\newblock CRC Press, Boca Raton, FL, 3rd edition, 2024.

\bibitem{wunderling2022}
N.~Wunderling, A.~Staal, B.~Sakschewski, M.~Hirota, O.~A. Tuinenburg, J.~F.
  Donges, H.~M.~J. Barbosa, and R.~Winkelmann.
\newblock Recurrent droughts increase risk of cascading tipping events by
  outpacing adaptive capacities in the {Amazon} rainforest.
\newblock {\em Proceedings of the National Academy of Sciences of the United
  States of America}, 119(32):e2120777119, 2022.

\bibitem{pastorsatorras2015}
R.~Pastor-Satorras, C.~Castellano, P.~Van~Mieghem, and A.~Vespignani.
\newblock Epidemic processes in complex networks.
\newblock {\em Reviews of Modern Physics}, 87(3):925--979, 2015.

\bibitem{scheffer2009}
M.~Scheffer, J.~Bascompte, W.~A. Brock, V.~Brovkin, S.~R. Carpenter, V.~Dakos,
  H.~Held, E.~H. van Nes, M.~Rietkerk, and G.~Sugihara.
\newblock Early-warning signals for critical transitions.
\newblock {\em Nature}, 461:53--59, 2009.

\bibitem{chen2012}
L.~Chen, R.~Liu, Z.-P. Liu, M.~Li, and K.~Aihara.
\newblock Detecting early-warning signals for sudden deterioration of complex
  diseases by dynamical network biomarkers.
\newblock {\em Scientific Reports}, 2:342, 2012.

\bibitem{dakos2012}
V.~Dakos, S.~R. Carpenter, W.~A. Brock, A.~M. Ellison, V.~Guttal, A.~R. Ives,
  S.~K\'efi, V.~Livina, D.~A. Seekell, E.~H. van Nes, and M.~Scheffer.
\newblock Methods for detecting early warnings of critical transitions in time
  series illustrated using simulated ecological data.
\newblock {\em {PloS ONE}}, 7(7):e41010, 2012.

\bibitem{kefi2014}
S.~K\'efi, V.~Guttal, W.~A. Brock, S.~R. Carpenter, A.~M. Ellison, V.~N.
  Livina, D.~A. Seekell, M.~Scheffer, E.~H. van Nes, and V.~Dakos.
\newblock Early warning signals of ecological transitions: Methods for spatial
  patterns.
\newblock {\em {PLoS ONE}}, 9(3):e92097, 2014.

\bibitem{speed2002}
T.~P. Speed.
\newblock John {W.} {Tukey's} contributions to analysis of variance.
\newblock {\em The Annals of Statistics}, 30(6):1649--1665, 2002.

\bibitem{cho2008}
E.~Cho and M.~J. Cho.
\newblock Variance of the with-replacement sample variance.
\newblock In {\em {JSM} Proceedings, Survey Research Methods Section}, pages
  1291--1293, Alexandria, VA, 2008. American Statistical Association.

\bibitem{vegassanchezferrero2012}
G.~Vegas-S\'anchez-Ferrero, S.~Aja-Fern\'andez, M.~Mart\'in-Fern\'andez, and
  C.~Palencia.
\newblock A direct calculation of moments of the sample variance.
\newblock {\em Mathematics and Computers in Simulation}, 82:790--804, 2012.

\bibitem{masuda2024}
N.~Masuda, K.~Aihara, and N.~G. MacLaren.
\newblock Anticipating regime shifts by mixing early warning signals from
  different nodes.
\newblock {\em Nature Communications}, 15:1086, 2024.

\bibitem{biggs2009}
R.~Biggs, S.~R. Carpenter, and W.~A Brock.
\newblock Turning back from the brink: Detecting an impending regime shift in
  time to avert it.
\newblock {\em Proceedings of the National Academy of Sciences of the United
  States of America}, 106(3):826--831, 2009.

\bibitem{dakos2010}
V.~Dakos, E.~H. van Nes, R.~Donangelo, H.~Fort, and M.~Scheffer.
\newblock Spatial correlation as leading indicator of catastrophic shifts.
\newblock {\em Theoretical Ecology}, 3:163--174, 2010.

\bibitem{burthe2016}
S.~J. Burthe, P.~A. Henrys, E.~B. Mackay, B.~M. Spears, R.~Campbell,
  L.~Carvalho, B.~Dudley, I.~D.~M. Gunn, D.~G. Johns, S.~C. Maberly, L.~May,
  M.~A. Newell, S.~Wanless, I.~J. Winfield, S.~J. Thackeray, and F.~Daunt.
\newblock Do early warning indicators consistently predict nonlinear change in
  long-term ecological data?
\newblock {\em Journal of Applied Ecology}, 53:666--676, 2016.

\bibitem{gsell2016}
A.~S. Gsell, U.~Scharfenberger, D.~\"Ozkundakci, A.~Walters, L.-A. Hansson,
  A.~B.~G. Janssen, P.~{N\~{o}ges}, P.~C. Reid, D.~E. Schindler, E.~Van Donk,
  V.~Dakos, and R.~Adrian.
\newblock Evaluating early-warning indicators of critical transitions in
  natural aquatic ecosystems.
\newblock {\em Proceedings of the National Academy of Sciences of the United
  States of America}, 113(50):E8089--E8095, 2016.

\bibitem{guttal2009}
V.~Guttal and C.~Jayaprakash.
\newblock Spatial variance and spatial skewness: Leading indicators of regime
  shifts in spatial ecological systems.
\newblock {\em Theoretical Ecology}, 2:3--12, 2009.

\bibitem{carpenter2010}
S.~R. Carpenter and W.~A. Brock.
\newblock Early warnings of regime shifts in spatial dynamics using the
  discrete {Fourier} transform.
\newblock {\em Ecosphere}, 1(5):art10, 2010.

\bibitem{dai2013}
L.~Dai, K.~S. Korolev, and J.~Gore.
\newblock Slower recovery in space before collapse of connected populations.
\newblock {\em Nature}, 496:355--358, 2013.

\bibitem{kefi2007}
S.~K\'efi, M.~Rietkerk, C.~L. Alados, Y.~Pueyo, V.~P. Papanastasis, A.~ElAich,
  and P.~C. de~Ruiter.
\newblock Spatial vegetation patterns and imminent desertification in
  {Mediterranean} arid ecosystems.
\newblock {\em Nature}, 449:213--217, 2007.

\bibitem{dakos2011}
V.~Dakos, S.~K\'efi, M.~Rietkerk, E.~H. van Nes, and M.~Scheffer.
\newblock Slowing down in spatially patterned ecosystems at the brink of
  collapse.
\newblock {\em The American Naturalist}, 177(6):E153--E166, 2011.

\bibitem{genin2018}
A.~G\'enin, S.~Majumder, S.~Sankaran, F.~D. Scheider, A.~Danet, M.~Berdugo,
  V.~Guttal, and S.~K\'efi.
\newblock Spatially heterogeneous stressors can alter the performance of
  indicators of regime shifts.
\newblock {\em Ecological Indicators}, 94:520--533, 2018.

\bibitem{sankaran2018}
S.~Sankaran, S.~Majumder, S.~K\'efi, and V.~Guttal.
\newblock Implications of being discrete and spatial for detecting early
  warning signals of regime shifts.
\newblock {\em Ecological Indicators}, 94:503--511, 2018.

\bibitem{ma2022}
Z.~Ma, Y.~Luo, C.~Zeng, and B.~Zheng.
\newblock Spatiotemporal diffusion as early warning signal for critical
  transitions in spatial tumor-immune system with stochasticity.
\newblock {\em Physical Review Research}, 4:023039, 2022.

\bibitem{genin2024}
A.~G\'enin, S.~A. Navarrete, A.~Garcia-Mayor, and E.~A. Wieters.
\newblock Emergent spatial patterns can indicate upcoming regime shifts in a
  realistic model of coral community.
\newblock {\em The American Naturalist}, 203(2):204--218, 2024.

\bibitem{donangelo2010}
R.~Donangelo, H.~Fort, V.~Dakos, M.~Scheffer, and E.~H. van Nes.
\newblock Early warnings for catastrophic shifs in ecosystems: Comparison
  between spatial and temporal indicators.
\newblock {\em International Journal of Bifurcation and Chaos}, 20(2):315--321,
  2010.

\bibitem{buelo2018}
C.~D. Buelo, S.~R. Carpenter, and M.~L. Pace.
\newblock A modeling analysis of spatial statistical indicators of thresholds
  for algal blooms.
\newblock {\em Limnology and Oceanography Letters}, 3:384--392, 2018.

\bibitem{majumder2019}
S.~Majumder, K.~Tamma, S.~Ramaswamy, and V.~Guttal.
\newblock Inferring critical thresholds of ecosystem transitions from spatial
  data.
\newblock {\em Ecology}, 100(7):e02722, 2019.

\bibitem{pal2022}
K.~Pal, S.~Deb, and P.~S. Dutta.
\newblock Tipping points in spatial ecosystems driven by short-range correlated
  noise.
\newblock {\em Physical Review E}, 106:054412, 2022.

\bibitem{wang2023}
J.~Wang, C.~Zeng, X.~Han, Z.~Ma, and B.~Zheng.
\newblock Detecting early warning signals of financial crisis in spatial
  endogenous credit model using patch-size distribution.
\newblock {\em Physica A}, 625:128925, 2023.

\bibitem{deb2024}
S.~Deb and P.~S. Dutta.
\newblock Critical transitions in spatial systems induced by
  {Ornstein--Uhlenbeck} noise: Spatial mutual information as a precursor.
\newblock {\em Proceedings of the Royal Academy of Sciences A}, 480:20230594,
  2024.

\bibitem{vandeleemput2014}
I.~A. van~de Leemput, M.~Wichers, A.~O.~J. Cramer, D.~Borsboom, F.~Tuerlinckx,
  P.~Kuppens, E.~H. van Nes, W.~Viechtbauer, E.~J. Giltay, S.~H. Aggen,
  C.~Derom, N.~Jacobs, K.~S. Kendler, H.~L.~J. van~der Maas, M.~C. Neale,
  F.~Peeters, E.~Thiery, P.~Zachar, and M.~Scheffer.
\newblock Critical slowing down as early warning for the onset and termination
  of depression.
\newblock {\em Proceedings of the National Academy of Sciences of the United
  States of America}, 111(1):87--92, 2014.

\bibitem{dablander2023}
F.~Dablander, A.~Pichler, A.~Cika, and A.~Bacilieri.
\newblock Anticipating critical transitions in psychological systems using
  early warning signals: Theoretical and practical considerations.
\newblock {\em Psychological Methods}, 28(4):765--790, 2023.

\bibitem{bascompte2023}
J.~Bascompte and M.~Scheffer.
\newblock The resilience of plant--pollinator networks.
\newblock {\em Annual Review of Entomology}, 68:363--380, 2023.

\bibitem{baguette2013}
M.~Baguette, S.~Blanchet, D.~Legrand, V.~M. Stevens, and C.~Turlure.
\newblock Individual dispersal, landscape connectivity and ecological networks.
\newblock {\em Biological Reviews}, 88:310--326, 2013.

\bibitem{white2018}
E.~R. White and A.~T. Smith.
\newblock The role of spatial structure in the collapse of regional
  metapopulations.
\newblock {\em Ecology}, 99(12):2815--2822, 2018.

\bibitem{litzow2008}
M.~A. Litzow, J.~D. Urban, and B.~J. Laurel.
\newblock Increased spatial variance accompanies reorganization of two
  continental shelf ecosystems.
\newblock {\em Ecological Applications}, 18(6):1331--1337, 2008.

\bibitem{cline2014}
T.~J. Cline, D.~A. Seekell, S.~R. Carpenter, M.~L. Pace, J.~R. Hodgson, J.~F.
  Kitchell, and B.~C. Weidel.
\newblock Early warnings of regime shifts: Evaluation of spatial indicators
  from a whole-ecosystem experiment.
\newblock {\em Ecosphere}, 5(8):102, 2014.

\bibitem{eby2017}
S.~Eby, A.~Agrawal, S.~Majumder, A.~P. Dobson, and V.~Guttal.
\newblock Alternative stable states and spatial indicators of critical slowing
  down along a spatial gradient in a savanna ecosystem.
\newblock {\em Global Ecology and Biogeography}, 26:638--649, 2017.

\bibitem{litzow2017}
M.~A. Litzow.
\newblock Indications of hysteresis and early warning signals of reduced
  community resilience during a {Bering Sea} cold anomaly.
\newblock {\em Marine Ecology Progress Series}, 571:13--28, 2017.

\bibitem{butitta2017}
V.~L. Butitta, S.~R. Carpenter, L.~C. Loken, M.~L. Pace, and E.~H. Stanley.
\newblock Spatial early warning signals in a lake manipulation.
\newblock {\em Ecosphere}, 8(10):e01941, 2017.

\bibitem{rindi2018}
L.~Rindi, M.~Dal Bello, and L.~Benedetti-Cecchi.
\newblock Experimental evidence of spatial signatures of approaching regime
  shifts in macroalgal canopies.
\newblock {\em Ecology}, 99(8):1709--1715, 2018.

\bibitem{tirabassi2023}
G.~Tirabassi and C.~Masoller.
\newblock Entropy-based early detection of critical transitions in spatial
  vegetation fields.
\newblock {\em Proceedings of the National Academy of Sciences of the United
  States of America}, 120(1):e2215667120, 2023.

\bibitem{rindi2024}
L.~Rindi, C.~Mintrone, C.~Ravaglioli, and L.~Benedetti-Cecchi.
\newblock Spatial signatures of an approaching regime shift in {{\em Posidonia
  oceanica}} meadows.
\newblock {\em Marine Environmental Research}, 198(2024):106499, 2024.

\bibitem{wang2024}
Y.~Wang, H.~Liu, W.~Zhao, J.~Jiang, Z.~He, Y.~Yu, L.~Guo, and O.~Yetemen.
\newblock Early warning signals of grassland ecosystem degredation: A case
  study from the northeast {Qinghai--Tibetan Plateau}.
\newblock {\em Catena}, 239:107970, 2024.

\bibitem{jentsch2018}
P.~C. Jentsch, M.~Anand, and C.~T. Bauch.
\newblock Spatial correlation as an early warning signal of regime shifts in a
  multiplex disease-behaviour network.
\newblock {\em Journal of Theoretical Biology}, 448:17--25, 2018.

\bibitem{phillips2020}
B.~Phillips, M.~Anand, and C.~T. Bauch.
\newblock Spatial early warning signals of social and epidemiological tipping
  points in a coupled behaviour-disease network.
\newblock {\em Scientific Reports}, 10:7611, 2020.

\bibitem{kouvaris2012}
N.~E. Kouvaris, H.~Kori, and A.~S. Mikhailov.
\newblock Traveling and pinned fronts in bistable reaction-diffusion systems on
  networks.
\newblock {\em {PLoS ONE}}, 7(9):e45029, 2012.

\bibitem{brummitt2015}
C.~D. Brummitt, G.~Barnett, and R.~M. D'Souza.
\newblock Coupled catastrophes: Sudden shifts cascade and hop among
  interdependent systems.
\newblock {\em Journal of The Royal Society Interface}, 12:20150712, 2015.

\bibitem{lever2020}
J.~J. Lever, I.~A. van~de Leemput, E.~Weinans, R.~Quax, V.~Dakos, E.~H. van
  Nes, J.~Bascompte, and M.~Scheffer.
\newblock Foreseeing the future of mutualistic communities beyond collapse.
\newblock {\em Ecology Letters}, 23:2--15, 2020.

\bibitem{gao2016}
J.~Gao, B.~Barzel, and A.-L. Barab{\'a}si.
\newblock Universal resilience patterns in complex networks.
\newblock {\em Nature}, 530:307--312, 2016.

\bibitem{legendre1989}
P.~Legendre and M.-J. Fortin.
\newblock Spatial pattern and ecological analysis.
\newblock {\em Vegetatio}, 80(2):107--138, 1989.

\bibitem{okabe2012}
A.~Okabe and K.~Sugihara.
\newblock {\em Spatial analysis along networks: Statistical and computational
  methods}.
\newblock John Wiley \& Sons, West Sussex, UK, 2012.

\bibitem{maclaren2023}
N.~G. MacLaren, P.~Kundu, and N.~Masuda.
\newblock Early warnings for multi-stage transitions in dynamics on networks.
\newblock {\em Journal of the Royal Society Interface}, 20:20220743, 2023.

\bibitem{boettiger2012}
C.~Boettiger and A.~Hastings.
\newblock Quantifying limits to detection of early warning for critical
  transitions.
\newblock {\em Journal of the Royal Society Interface}, 9:2527--2539, 2012.

\bibitem{chen2022}
S.~Chen, A.~Ghadami, and B.~I. Epureanu.
\newblock Practical guide to using {Kendall's} $\tau$ in the context of
  forecasting critical transitions.
\newblock {\em Royal Society Open Science}, 9:211346, 2022.

\bibitem{xu2015}
C.~Xu, E.~H. van Nes, M.~Holmgren, S.~K\'efi, and M.~Scheffer.
\newblock Local facilitation may cause tipping points on a landscape level
  preceded by early-warning indicators.
\newblock {\em The American Naturalist}, 186(4):E81--E90, 2015.

\bibitem{tirabassi2024}
G.~Tirabassi.
\newblock Linear theory of the spatial signatures of critical slowing down.
\newblock {\em Physical Review Research}, 6:023228, 2024.

\bibitem{barzel2013}
B.~Barzel and A.-L. Barab{\'a}si.
\newblock Universality in network dynamics.
\newblock {\em Nature Physics}, 9:673--681, 2013.

\bibitem{hens2019}
C.~Hens, U.~Harush, S.~Haber, R.~Cohen, and B.~Barzel.
\newblock Spatiotemporal signal propagation in complex networks.
\newblock {\em Nature Physics}, 15(4):403--412, 2019.

\bibitem{barzon2024}
G.~Barzon, O.~Artime, S.~Suweis, and M.~De Domenico.
\newblock Unraveling the mesoscale organization induced by network-driven
  processes.
\newblock {\em Proceedings of the National Academy of Sciences of the United
  States of America}, 121(28):e2317608121, 2024.

\bibitem{boettiger2013}
C.~Boettiger, N.~Ross, and A.~Hastings.
\newblock Early warning signals: The charted and uncharted territories.
\newblock {\em Theoretical Ecology}, 6:255--264, 2013.

\bibitem{bury2021}
T.~M. Bury, R.~I. Sujith, I.~Pavithran, M.~Scheffer, T.~M. Lenton, M.~Anand,
  and C.~T. Bauch.
\newblock Deep learning for early warning signals of tipping points.
\newblock {\em Proceedings of the National Academy of Sciences of the United
  States of America}, 118(39):e2106140118, 2021.

\bibitem{obrien2023}
D.~A. {O'Brien}, S.~Deb, G.~Gal, S.~J. Thackeray, P.~S. Dutta, S.~S. Matsuzaki,
  L.~May, and C.~F. Clements.
\newblock Early warning signals have limited applicability to empirical lake
  data.
\newblock {\em Nature Communications}, 14:7942, 2023.

\bibitem{southall2021}
E.~Southall, T.~S. Brett, M.~J. Tildesley, and L.~Dyson.
\newblock Early warning signals of infectious disease transitions: A review.
\newblock {\em Journal of the Royal Society Interface}, 18:20210555, 2021.

\bibitem{KONECT}
J.~Kunegis.
\newblock {KONECT} -- {The} {Koblenz} {Network} {Collection}.
\newblock In {\em Proceedings of the 22nd International Conference on the World
  Wide Web}, pages 1343--1350, 2013.

\bibitem{Netzschleuder}
T.~P. Peixoto.
\newblock The {Netzschleuder} network catalogue and repository, 2020.
\newblock \url{https://networks.skewed.de/}, accessed 18 April 2024.

\bibitem{igraph}
G.~Cs\'ardi, T.~Nepusz, V.~Traag, Sz. Horv\'at, F.~Zanini, D.~Noom, and
  K.~M\"uller.
\newblock {\em {igraph}: {Network} Analysis and Visualization in {R}}, 2024.
\newblock R package version 2.0.3.

\bibitem{networkx}
A.~A. Hagberg, D.~A. Schult, and P.~J. Swart.
\newblock Exploring network structure, dynamics, and function using {NetworkX}.
\newblock In G.~Varoquaux, T.~Vaught, and J.~Millman, editors, {\em Proceedings
  of the 7th Python in Science Conference}, pages 11--15, Pasadena, CA USA,
  2008.

\bibitem{descormiers2011}
K.~Descormiers and C.~Morselli.
\newblock Alliances, conflicts, and contradictions in {Montreal's} street gang
  landscape.
\newblock {\em International Criminal Justice Review}, 21(3):297--314, 2011.

\bibitem{baird1989}
D.~Baird and R.~E. Ulanowicz.
\newblock The seasonal dynamics of the {Chesapeake Bay} ecosystem.
\newblock {\em Ecological Monographs}, 59(4):329--364, 1989.

\bibitem{freeman1988}
L.~C. Freeman, S.~C. Freeman, and A.~G. Michaelson.
\newblock On human social intelligence.
\newblock {\em Journal of Social and Biological Structures}, 11(4):415--425,
  1988.

\bibitem{knuth2008}
D.~E. Knuth.
\newblock {\em The Art of Computer Programming, Volume 4, Fascicle 0:
  Introduction to Combinatorial and Boolean Functions}.
\newblock Addison-Wesley, Upper Saddle River, NJ, 2008.

\bibitem{thompson2003}
R.~M. Thompson and C.~R. Townsend.
\newblock Impacts on stream food webs of native and exotic forest: {An}
  intercontinental comparison.
\newblock {\em Ecology}, 84(1):145--161, 2003.

\bibitem{lusseau2003}
D.~Lusseau, K.~Schneider, O.~J. Boisseau, P.~Haase, E.~Slooten, and S.~M.
  Dawson.
\newblock The bottlenose dolphin community of {Doubtful Sound} features a large
  proportion of long-lasting associations.
\newblock {\em Behavioral Ecology and Sociobiology}, 54:396--405, 2003.

\bibitem{hayes2006}
B.~Hayes.
\newblock Connecting the dots.
\newblock {\em American Scientist}, 94(5):400--404, 2006.

\bibitem{correia2019}
R.~B. Correia, L.~P. de~Ara{\'u}jo~Kohler, M.~M. Mattos, and L.~M. Rocha.
\newblock City-wide electronic health records reveal gender and age biases in
  administration of known drug--drug interactions.
\newblock {\em {NPJ} Digital Medicine}, 2:74, 2019.

\bibitem{haraldsdottir1992}
S.~Haraldsdottir, S.~Gupta, and R.~M. Anderson.
\newblock Preliminary studies of sexual networks in a male homosexual community
  in iceland.
\newblock {\em Journal of Acquired Immune Deficiency Syndromes}, 5:374--381,
  1992.

\bibitem{hoglund2006}
M.~H\"oglund, A.~Frigyesi, and F.~Mitelman.
\newblock A gene fusion network in human neoplasia.
\newblock {\em Oncogene}, 25(18):2674--2678, 2006.

\bibitem{newman2006}
M.~E.~J. Newman.
\newblock Finding community structure in networks using the eigenvectors of
  matrices.
\newblock {\em Physical Review E}, 74(3):036104, 2006.

\bibitem{girvan2002}
M.~Girvan and M.~E.~J. Newman.
\newblock Community structure in social and biological networks.
\newblock {\em Proceedings of the National Academy of Sciences of the United
  States of America}, 99(12):7821--7826, 2002.

\bibitem{coleman1957}
J.~Coleman, E.~Katz, and H.~Menzel.
\newblock The diffusion of an innovation among physicians.
\newblock {\em Sociometry}, 20(4):253--270, 1957.

\bibitem{fire2012}
M.~Fire, G.~Katz, Y.~Elovici, B.~Shapira, and L.~Rokach.
\newblock Predicting student exam's scores by analyzing social network data.
\newblock In R.~Huang, A.~A. Ghorbani, G.~Pasi, T.~Yamaguchi, N.~Y. Yen, and
  B.~Jin, editors, {\em 8th International Conference on Active Media
  Technology}, pages 584--595, 2012.

\bibitem{beuming2005}
T.~Beuming, L.~Skrabanek, M.~Y. Niv, P.~Mukherjee, and H.~Weinstein.
\newblock {PDZBase}: A protein–protein interaction database for
  {PDZ}-domains.
\newblock {\em Bioinformatics}, 21(6):827--828, 2005.

\bibitem{michalski2011}
R.~Michalski, S.~Palus, and P.~Kazienko.
\newblock Matching organizational structure and social network extracted from
  email communication.
\newblock In W.~Abramowicz, editor, {\em 14th International Conference on
  Business Information Systems}, volume~87 of {\em Lecture Notes in Business
  Information Processing}, pages 197--206, 2011.

\bibitem{chami2017}
G.~F. Chami, S.~E. Ahnert, N.~B. Kabatereine, and E.~M. Tukahebwa.
\newblock Social network fragmentation and community health.
\newblock {\em Proceedings of the National Academy of Sciences of the United
  States of America}, 114(36):E7425--E7431, 2017.

\bibitem{gleiser2003}
P.~M. Gleiser and L.~Danon.
\newblock Community structure in jazz.
\newblock {\em Advances in Complex Systems}, 6(4):565--573, 2003.

\bibitem{subelj2011b}
L.~\v{S}ubelj and M.~Bajec.
\newblock Community structure of complex software systems: Analysis and
  applications.
\newblock {\em Physica A}, 390(16):2968--2975, 2011.

\bibitem{shenorr2002}
S.~S. Shen-Orr, R.~Milo, S.~Mangan, and U.~Alon.
\newblock Network motifs in the transcriptional regulation network of {{\em
  Escherichia coli}}.
\newblock {\em Nature Genetics}, 31:64--68, 2002.

\bibitem{dedomenico2014}
M.~De Domenico, A.~Sol\'e-Ribalta, S.~G\'omez, and A.~Arenas.
\newblock Navigability of interconnected networks under random failures.
\newblock {\em Proceedings of the National Academy of Sciences of the United
  States of America}, 111(23):8351--8356, 2014.

\bibitem{sun2016}
J.~Sun, J.~Kunegis, and S.~Staab.
\newblock Predicting user roles in social networks using transfer learning with
  feature transformation.
\newblock In C.~Domeniconi, F.~Gullo, F.~Bonchi, J.~Domingo-Ferrer,
  R.~Baeza-Yates, Z.-H. Zhou, and X.~Wu, editors, {\em 2016 IEEE 16th
  International Conference on Data Mining Workshops}, pages 128--135, 2016.

\bibitem{isella2011}
L.~Isella, J.~Stehl{\'e}, A.~Barrat, C.~Cattuto, J.-F. Pinton, and W.~Van~den
  Broeck.
\newblock What's in a crowd? {Analysis} of face-to-face behavioral networks.
\newblock {\em Journal of Theoretical Biology}, 271:166--180, 2011.

\bibitem{jeong2000}
H.~Jeong, B.~Tombor, R.~Albert, Z.~N. Oltvai, and A.-L. Barab{\'a}si.
\newblock The large-scale organization of metabolic networks.
\newblock {\em Nature}, 407:651--654, 2000.

\bibitem{cook2019}
S.~J. Cook, T.~A. Jarrell, C.~A. Brittin, Y.~Wang, A.~E. Bloniarz, M.~A.
  Yakovlev, K.~C.~Q. Nguyen, L.~T.-H. Tang, E.~A. Bayer, J.~S. Duerr, H.~E.
  B\"ulow, O.~Hobert, D.~H. Hall, and S.~W. Emmons.
\newblock Whole-animal connectomes of both {{\em Caenorhabditis elegans}}
  sexes.
\newblock {\em Nature}, 571:63--71, 2019.

\bibitem{milo2002}
R.~Milo, S.~Shen-Orr, S.~Itzkovitz, N.~Kashtan, D.~Chklovskii, and U.~Alon.
\newblock Network motifs: Simple building blocks of complex networks.
\newblock {\em Science}, 298(5594):824--827, 2002.

\bibitem{hidalgo2007}
C.~A. Hidalgo, B.~Klinger, A.-L. Barab\'asi, and R.~Hausmann.
\newblock The product space conditions the development of nations.
\newblock {\em Science}, 317:482--487, 2007.

\bibitem{watts1998}
D.~J. Watts and S.~H. Strogatz.
\newblock Collective dynamics of ``small-world'' networks.
\newblock {\em Nature}, 393:440--442, 1998.

\bibitem{barabasi1999}
A.-L. Barab{\'a}si and R.~Albert.
\newblock Emergence of scaling in random networks.
\newblock {\em Science}, 286(5439):509--512, 1999.

\bibitem{holme2002}
P.~Holme and B.~J. Kim.
\newblock Growing scale-free networks with tunable clustering.
\newblock {\em Physical Review E}, 65(2):026107, 2002.

\bibitem{goh2001}
K.-I. Goh, B.~Kahng, and D.~Kim.
\newblock Universal behavior of load distribution in scale-free networks.
\newblock {\em Physical Review Letters}, 87(27):278701, 2001.

\bibitem{chung2002}
F.~Chung and L.~Lu.
\newblock Connected components in random graphs with given expected degree
  sequences.
\newblock {\em Annals of Combinatorics}, 6:125--145, 2002.

\bibitem{cho2009}
Y.~S. Cho, J.~S. Kim, J.~Park, B.~Kahng, and D.~Kim.
\newblock Percolation transitions in scale-free networks under the {Achlioptas}
  process.
\newblock {\em Physical Review Letters}, 103(13):135702, 2009.

\end{thebibliography}

\appendix
\setcounter{section}{0}
\setcounter{figure}{0}
\setcounter{table}{0}
\setcounter{equation}{0}

\renewcommand{\thesection}{S\arabic{section}}
\renewcommand{\thefigure}{S\arabic{figure}}
\renewcommand{\thetable}{S\arabic{table}}
\renewcommand{\theequation}{S\arabic{equation}}

\renewcommand*{\theHsection}{S.\thesection}
\renewcommand*{\theHfigure}{S.\thefigure}
\renewcommand*{\theHtable}{S.\thetable}
\renewcommand*{\theHequation}{S.\theequation}

\setlength{\tabcolsep}{10pt}

\begin{center}
  \vspace*{12pt}
  {\large \bf Supplementary Materials for:\\
    \vspace{12pt} Applicability of spatial early warning signals to complex network dynamics}
  \vspace{12pt} \\
  Neil G. MacLaren, Kazuyuki Aihara, and Naoki Masuda
\end{center}

\section{Networks\label{sec:SInetworks}}

We used 35 networks, i.e., 30 empirical networks and five random synthetic networks. We also separately used one deterministic network, i.e., the periodic two-dimensional lattice. We used the largest connected component of each network, removed multiple edges, and coerced each network to be undirected and unweighted if not originally so. Table \ref{tab:networks} has descriptive information on each network. We downloaded the empirical networks from either the KONECT \cite{KONECT} or Netzschleuder \cite{Netzschleuder} repository. We generated model networks with igraph \cite{igraph} except the Holme-Kim (HK) network, which we generated with NetworkX \cite{networkx}.

\afterpage{
  % \begin{longtable}{p{0.8in}p{0.11in}p{0.11in}p{4.4in}}
  \begin{longtblr}[
    caption = {Networks used in this study. $N$: number of nodes, $M$: number of edges.},
    label = {tab:networks}
    ]{colspec = {X[1.5, l] X[0.5, l] X[0.5, l] X[6, l]}}
    % \caption{
    %   Networks used in this study. $N$: number of nodes, $M$: number of edges.
    % }
    % \label{tab:networks}\\
    \toprule
    Network name & $N$ & $M$ & Notes \\
    \midrule
    Montreal & 29 & 75 & A network of relationships between gangs in Montreal, Quebec \cite{descormiers2011}.\\
    Chesapeake & 39 & 170 & A saltwater trophic network in which the nodes are major ecosystem components, such as phytoplankton or fish larvae. Edges are carbon flows between them \cite{baird1989}.\\
    Windsurfer & 43 & 336 & A network of interpersonal contacts between windsurfers \cite{freeman1988}.\\
    Geographic & 49 & 107 & A network of neighboring US states and territories \cite{knuth2008}.\\
    Catlins & 59 & 110 & A freshwater trophic network \cite{thompson2003}. Nodes are organism taxa. Edges record which taxa were found to consume which other taxa.\\
    Dolphin & 62 & 159 & A social network of wild dolphins \cite{lusseau2003}.\\
    Terrorist & 64 & 243 & A network of contacts between individuals involved in the train bombing in 2004 in Madrid, Spain \cite{hayes2006}.\\
    Drug interaction & 75 & 181 & A network of drug interactions in the health records of individuals in Blumenau, Brazil \cite{correia2019}.\\
    Contact & 75 & 114 & A network of sexual contacts among Icelandic individuals \cite{haraldsdottir1992}.\\
    Canton & 109 & 717 & A freshwater trophic network \cite{thompson2003}. See ``Catlins'' above.\\
    Gene fusion & 110 & 124 & A network of genes which have been observed to have fused in human neoplasia \cite{hoglund2006}.\\
    Word & 112 & 425 & A network of word co-occurrence \cite{newman2006}.\\
    Football & 115 & 613 & A network of US college American football games \cite{girvan2002}. Nodes are US collegiate football teams and are adjacent if the two teams played a game during the 2000 season.\\
    Physician & 117 & 465 & A social network of physicians \cite{coleman1957}.\\
    Student & 141 & 256 & A cooperation network of university students \cite{fire2012}.\\
    Protein & 161 & 209 & A protein interaction network \cite{beuming2005}.\\
    Email & 167 & 3250 & A network of emails from a manufacturing company \cite{michalski2011}. Nodes are email accounts and are adjacent if an email was exchanged between them.\\
    Village & 187 & 431 & An advice network from an Ugandan village \cite{chami2017}. Nodes are households and are adjacent if an individual from one household nominated an individual from another as a source of advice.\\
    Jazz player & 198 & 2742 & A collaboration network of jazz musicians \cite{gleiser2003}. Nodes are musicians and are adjacent if the two musicians played in an ensemble together.\\
    Flamingo software & 228 & 491 & A software dependency network \cite{subelj2011b}. Each node is a class, in the sense of object oriented programming. Edges represent the existence of a dependency between two classes.\\
    {\em E. coli} & 328 & 456 & A transcription network of the bacterium {\em Escherichia coli} \cite{shenorr2002}. Nodes are operons (gene clusters). Edges represent a regulatory relationship between the two operons.\\
    Transportation & 369 & 430 & A network of train stations in London, UK \cite{dedomenico2014}.\\
    Coauthorship & 379 & 914 & A coauthorship network \cite{newman2006}. Nodes are researchers working in network science. Edges indicate that two researchers coauthored a paper.\\ 
    Wikipedia user & 404 & 734 & A network of users of the Haitian Creole Wikipedia page \cite{sun2016}. Two nodes are adjacent if one of the two users wrote on the other's talk page.\\
    Proximity & 410 & 2765 & A network of face-to-face contacts at a museum display \cite{isella2011}. Nodes represent museum visitors and are adjacent if the two visitors were sufficiently close to each other in physical space during their visit.\\
    {\em C. elegans}: metabolic & 453 & 2025 & A metabolic network of the nematode {\em Caenorhabditis elegans} \cite{jeong2000}. Nodes are enzymes, substrates, or temporary complexes and are adjacent if they are involved in a chemical reaction together.\\ 
    {\em C. elegans}: neuronal & 460 & 1432 & A neuronal network of {\em C. elegans} \cite{cook2019}. Nodes are neurons. Edges represent synapses.\\
    {\em S. cerevisiae} & 664 & 1065 & A network of operons in the yeast {\em Saccharomyces cerevisiae} \cite{milo2002}. See ``{\em E. coli}'' above.\\
    Product & 774 & 1779 & A network of exported products \cite{hidalgo2007}. Nodes are economic products and are adjacent if they are sufficiently similar in terms of the quantities exported by the same countries.\\
    Jung software & 879 & 2047 & A software dependency network \cite{subelj2011b}. See the description for ``Flamingo'' above.\\
    ER & 100 & 249 & An Erd\H{o}s-R\'enyi random graph with connection probability $p=0.05$. \\
    WS & 100 & 400 & A Watts-Strogatz small-world random graph \cite{watts1998}. The seed graph is a one-dimensional periodic lattice with each node connected to four nearest neighbors. The rewiring probability is $p=0.02$. \\
    BA & 100 & 197 & A Barab\'asi-Albert random graph with $m=2$ \cite{barabasi1999}. The seed graph is a complete graph with $N=3$.\\
    HK & 100 & 196 & A Holme-Kim random graph with $m=2$ and an average local clustering coefficient of 0.22 \cite{holme2002}. The seed graph is an empty graph with $N=m$.\\
    GKK & 96 & 300 & A Goh-Kahng-Kim random graph \cite{goh2001}. Each edge $(i, j)$, $i, j \in \{1, \ldots N\}$, is present with probability $\frac{f_if_j}{(\sum_{\ell=1}^N f_\ell)^2}$, where $f_i = (i + i_0 -1)^{-\alpha}$, and $i_0 = N^{1-\frac{1}{\alpha}}\left[ 10\sqrt{2}(1 - \alpha) \right]^{\frac{1}{\alpha}}$ constrains the maximum degree \cite{chung2002, cho2009}. We set $\alpha=1$, $N=100$, and $M=300$ and took the largest connected component of the resulting graph.\\
    Lattice & 100 & 200 & A two-dimensional lattice with linear length 10 and periodic boundary conditions.\\
    \bottomrule
  % \end{longtable}
  \end{longtblr}
}

\clearpage
\newpage

\section{Simulation details\label{sec:SIsimulations}}

We list the parameters for each simulation condition and network in the separate SI file, ``Simulation parameters.ods.'' Specifically, for each simulation condition and network, we list (i) the control parameter values that define the simulation range, (ii) the value of the other control parameter (i.e., $u$ if the control parameter is $D$ and vice versa), which is fixed for each simulation sequence, and (iii) the value of $\Delta t$.

We list the values of the simulation range of the control parameter in columns G and H. For example, the first row shows the simulation parameters for ascending simulations of the coupled double-well dynamics with $D$ as the control parameter on the BA network. In this case, we established through trial and error that the simulation range $D \in [0, 0.15]$ satisfied the required properties stated in the main text (i.e., a well-defined simulation range and a sufficient number of control parameter values in the home range). When $D=0$, all nodes are far from the tipping point; when $D=0.15$, some nodes are near the tipping point and at least one node has already transitioned to the alternate state. Among other factors, the number of nodes and the degree distribution of the network affect distributions of the $x_i^*$ values in general, even if the simulation condition is the same, resulting in differences in the control parameter value at the tipping point. For example, the next row shows that $D=0.025$ is sufficient to observe a tipping point for the larger ``{\em C. elegans}: metabolic'' network. In row 7, as another example, we used $D=0.2$ for the smaller ``Catlins'' network to meet the same requirements.

To induce a bifurcation with one control parameter, we set the other parameter to a fixed value. We typically set $u=0$ if the control parameter was $D$. For descending simulations of the coupled double-well and mutualistic species dynamics, no node will undergo a transition while $D$ is non-negative. In those conditions, we set $u=-5$. If the control parameter was $u$, we typically set $D=0.05$ except for the gene-regulatory dynamics, for which we set $D=1$. We have adjusted these fixed values to support our requirements and show these values in columns C and D. For example, we show on line 76 that we used $D=0.005$ in ascending simulations of the coupled double-well dynamics with $u$ as the control parameter on the ``{\em C. elegans}: metabolic'' network. This network has a large maximum degree, leading to bifurcations at small values of $D$ (see line 4, which shows the simulation range when $D$ is the control parameter, $[0, 0.025]$).

Finally, we use $\Delta t = 0.01$ for most simulations. Some combinations of simulation condition and network led to numerical instability in our quadrature algorithm. In those cases, we set $\Delta t < 0.01$ and list the value of $\Delta t$ in column I.

\clearpage

\section{Comparison of $\tau'$ among the different early warning signals\label{sec:SItau}}

\begin{figure}[b]
  \includegraphics[width = 0.95\textwidth]{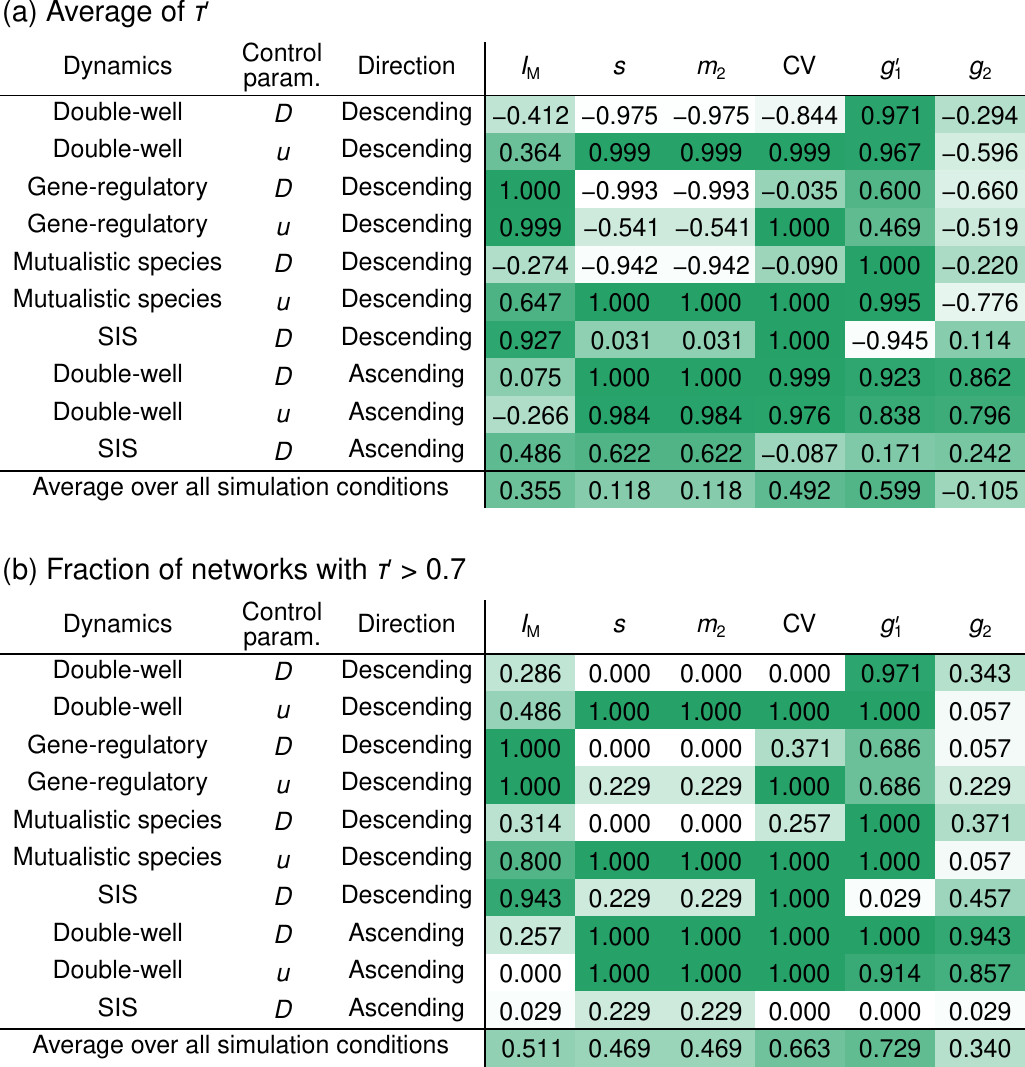}
  \caption{Performance of the six EWSs in terms of $\tau'$. (a) Average of $\tau'$ over the 35 networks.
  (b) Fraction over the 35 networks yielding $\tau' > 0.7$. A darker cell indicates a higher performance. ``Control param.'' abbreviates control parameter. For $I_{\text{M}}$, $\text{CV}$, $g_1'$, and $g_2$, these results are a summary of those for the individual networks shown in Fig.~3 in the main text.}
  \label{fig:tau-table}
\end{figure}

For each EWS and simulation condition, we show the average $\tau'$ values over the 35 networks and the fraction of networks among the 35 networks for which $\tau' > 0.7$ in Fig.~\ref{fig:tau-table}a and b, respectively. Two main findings are as follows. First, overall, i.e., as the average over all the simulation conditions (see the last row in each panel of the figure), $g_1'$ performs the best, $\text{CV}$ the second, and $I_{\text{M}}$ the third according to both criteria. Second, $\text{CV}$ performs the best among the three variants of the sample variance, i.e., $s$, $m_2$, and $\text{CV}$. The latter is because of the perfect performance of $\text{CV}$ under two simulation conditions, where $s$ and $m_2$ perform poorly (i.e., (gene-regulatory, $u$, descending) and (SIS, $D$, descending)). Although $s$ and $m_2$ perform substantially better than $\text{CV}$ under one simulation condition (i.e., (SIS, $D$, ascending)), the performance of $s$ and $m_2$ is not satisfactory, with only 22.9\% of networks yielding $\tau' > 0.7$.

\clearpage
\newpage

\section{Robustness with respect to the initial value of the control parameter\label{sec:robustness}}

Our classification scheme for EWSs depends on the initial value of the control parameter (i.e., $D$ or $u$); the final control parameter value is the one just before the tipping event. Therefore, to assess robustness of our results, we assessed the performance of the EWSs when we shifted the initial value of the control parameter 25\% or 50\% towards its final value within the home range.

\begin{figure}[b]
  \includegraphics[width = 0.95\textwidth]{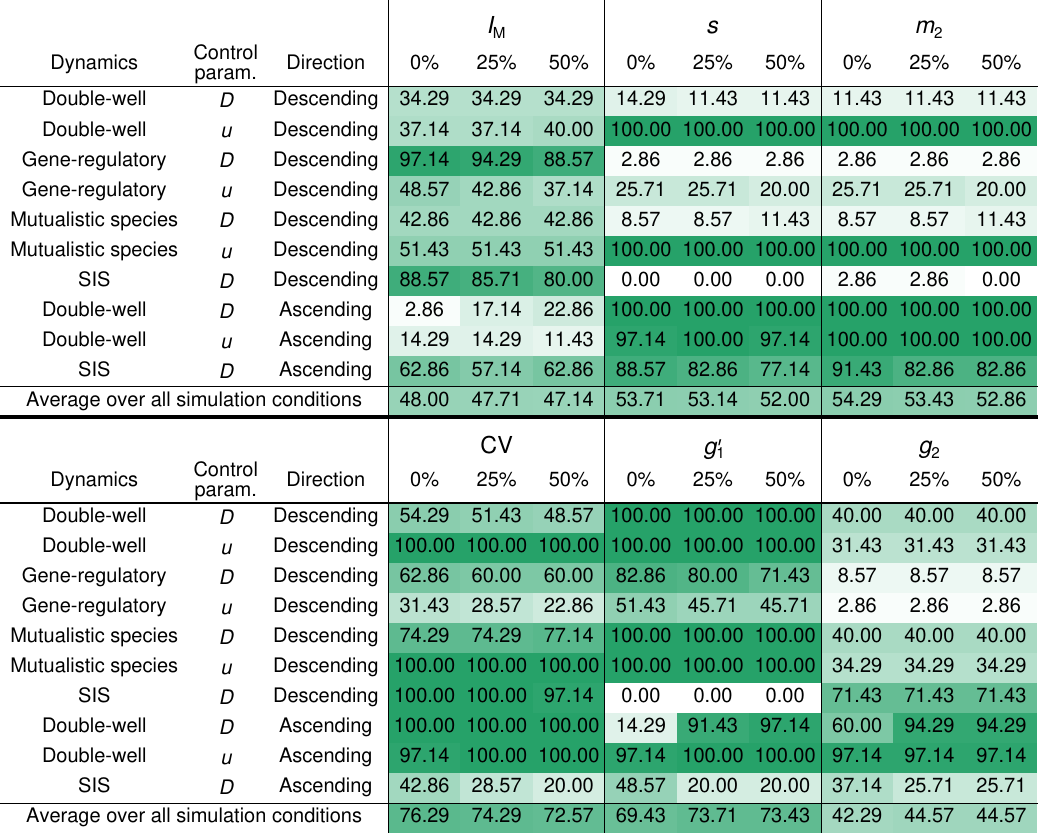}
  \caption{Performance of the EWSs for three initial values of the control parameter. The values shown are the percentage of the networks among the 35 networks for which the EWS is successful. Each numerical result cell corresponds to a simulation condition. A darker cell indicates a higher performance. ``Control param.'' abbreviates control parameter. In the last nine columns, 0\%, 25\%, and 50\% represent how much we shift the initial value of the control parameter towards its final value. Note that 0\% means the original home range. Therefore, the results for $I_{\text{M}}$, $\text{CV}$, $g_1'$, and $g_2$ with 0\% replicate those shown in Fig.~4 in the main text.}
  \label{fig:25-50}
\end{figure}

We show the classification results for the 25\% and 50\% cases as well as the original case (i.e., 0\% shift) in Fig.~\ref{fig:25-50}. The percentages shown in the figure indicate the fraction of networks among our 35 networks for which the EWS is successful. We find that the results are similar among the three values of the initial control parameter. There are only five out of 60 combinations of the simulation condition and the EWS for which the performance among any of the 0\%, 25\%, and 50\% cases is more than 20\% different, i.e.,
(double-well, $D$, ascending, $I_{\text{M}}$),
(SIS, $D$, ascending, $\text{CV}$),
(double-well, $D$, ascending, $g_1^{\prime}$),
(SIS, $D$, ascending, $g_1^{\prime}$), and
(double-well, $D$, ascending, $g_2$).

\clearpage
\newpage

\section{Behavior of $g_1'$ on the square lattice\label{sec:g1'-square-lattice}}

\begin{figure}[b]
  \includegraphics[width = 1.0\textwidth]{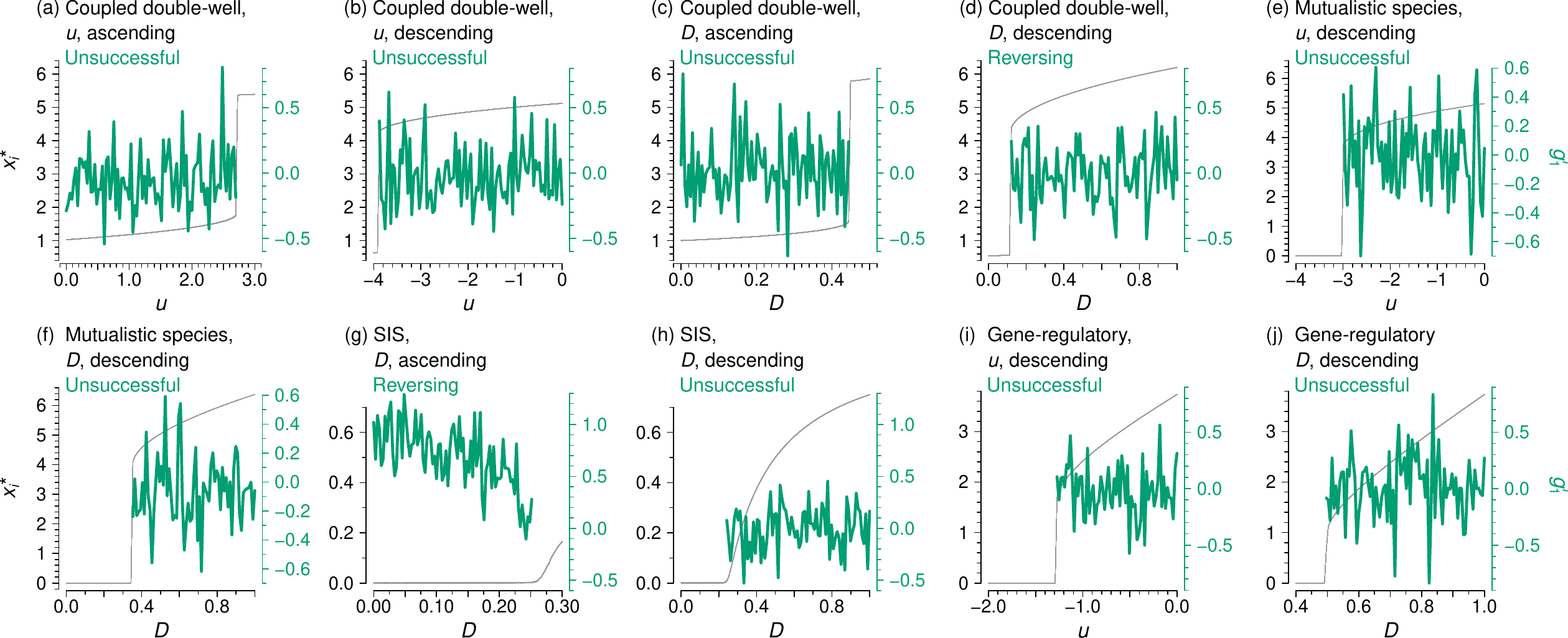}
  \caption{Spatial skewness, $g_1'$, on the square lattice as a function of the control parameter value. The green lines represent $g_1'$. The gray lines represent $x_i^*$.}
  \label{fig:skewness-lattice}
\end{figure}

We show the behavior of $g_1'$ on the square lattice by the green lines in Fig.~\ref{fig:skewness-lattice}. Each panel corresponds to one of the ten simulation conditions. As we mentioned in the main text, $g_1'$ is successful in two out of the ten simulation conditions, as indicated in the figure. In the eight simulation conditions under which $g_1'$ is unsuccessful (i.e.,  Fig.~\ref{fig:skewness-lattice}b--i), we find that $g_1'$ fluctuates around $0$, despite with a large variation, as the control parameter varies. Similar behavior of $g_1'$ is also present in Fig.~\ref{fig:skewness-lattice}d, in which $g_1'$ is classified as successful (precisely speaking, ``reversing''). However, given that $g_1'$ is high fluctuating as $D$ varies, we regard that this classification result is a false positive.

\clearpage
\newpage

\section{Examples of the behavior of spatial early warning signals on the square lattice\label{sec:SIlattice}}
 
\begin{figure}[b]
  \includegraphics[width = 0.7\textwidth]{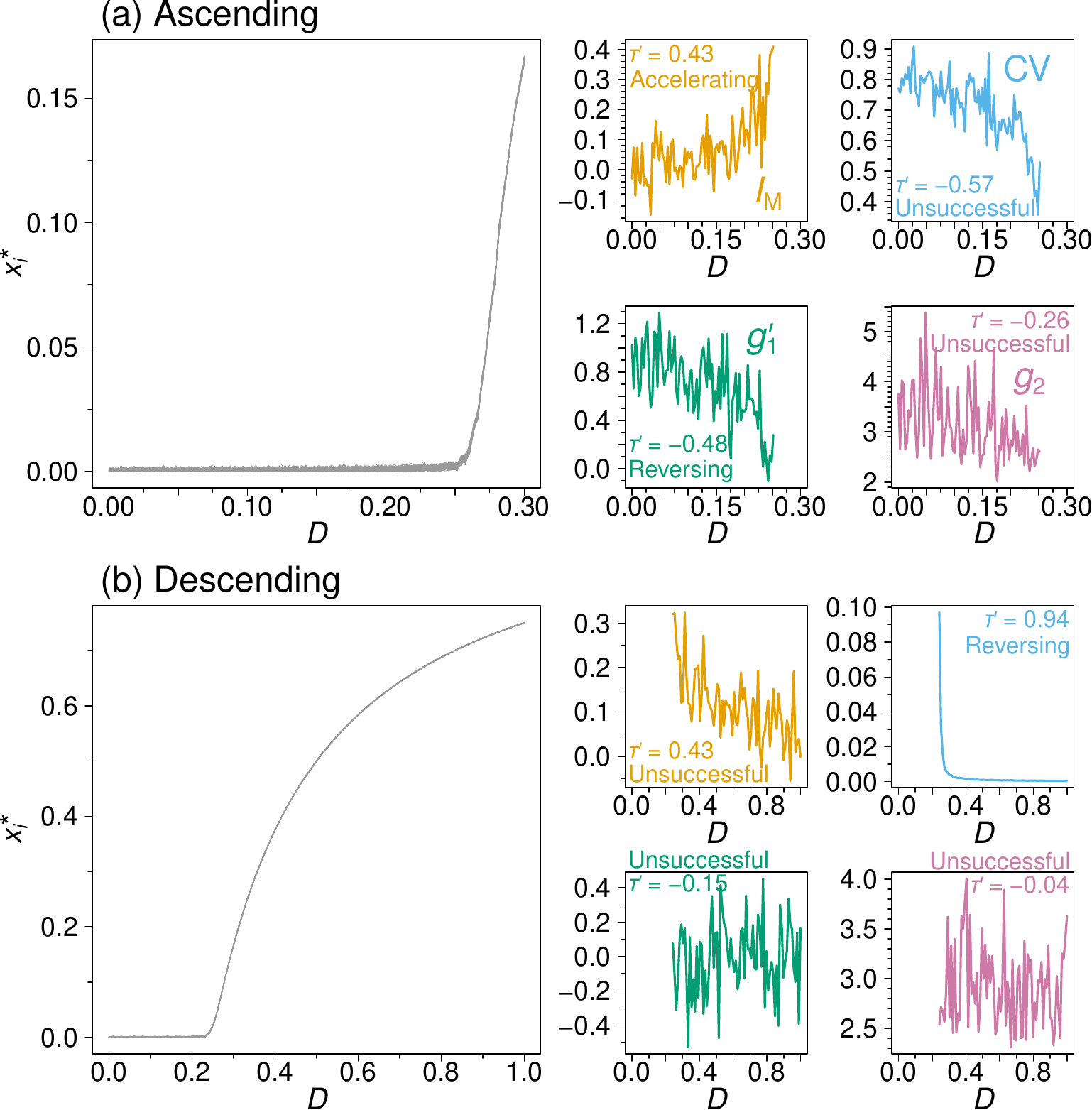}
  \caption{
    Node states and EWSs as a function of $D$ for the SIS dynamics on the square lattice.
    (a) Ascending simulations. (b) Descending simulations.
  }
  \label{fig:example-lattice}
\end{figure}

In the main text, we noted that, except for $\text{CV}$, spatial EWSs are noisier on the lattice network, even when they are classified as successful. We show in Fig.~\ref{fig:example-lattice} two examples of noisy behavior of the EWSs. Figure \ref{fig:example-lattice} is similar to Fig.~2 in the main text, except that we show ascending (Fig.~\ref{fig:example-lattice}a) and descending (Fig.~\ref{fig:example-lattice}b) simulations of the SIS dynamics on the square lattice.

\clearpage

\section{Regime shift without critical slowing down\label{sec:no-csd}}

We investigated the behavior of our spatial EWSs when regime shifts occur without critical slowing down. Specifically, we consider two such scenarios considered in \cite{boettiger2012, boettiger2013}.
To this end, we simulated the coupled double-well dynamics on two networks, the Catlins and Village networks, as follows. The simulation setups are similar to those in Supplementary Note 7 of \cite{masuda2024}.

The first scenario is when dynamical noise induces large jumps in $x_i(t)$. We poise the coupled double-well dynamics near the saddle-node bifurcation by setting $u = -3$, $D = 0.05$, and $\sigma = 0.25$ for the Catlins network and $\sigma = 0.2$ for the Village network. The other parameter values are the same as those used in the main text. We set $x_{i, 0} = 5$, $\forall i$. Then, we run a single 50 time-unit-long simulation for each network. In each of the two networks, we observe two nodes which, by random chance, transition from their upper to the lower state. Note that different simulation runs may result in more or fewer such node transitions. We computed $I_{\rm M}$, CV, $g_1'$, and $g_2$ from $t=5$, discarding initial transients, until the first node transitioned to the lower state (defined as $x_i(t) < 3$) using one sample from $x_i(t)$, $\forall i$ per time unit. Because $t$ is always increasing, a desirable EWS should have a positive $\tau$ value; although the system begins in the upper state and transits to the lower state, there is no steadily decreasing control parameter causing a decrease in node states.

\begin{figure}[b]
  \includegraphics[width = 0.7\textwidth]{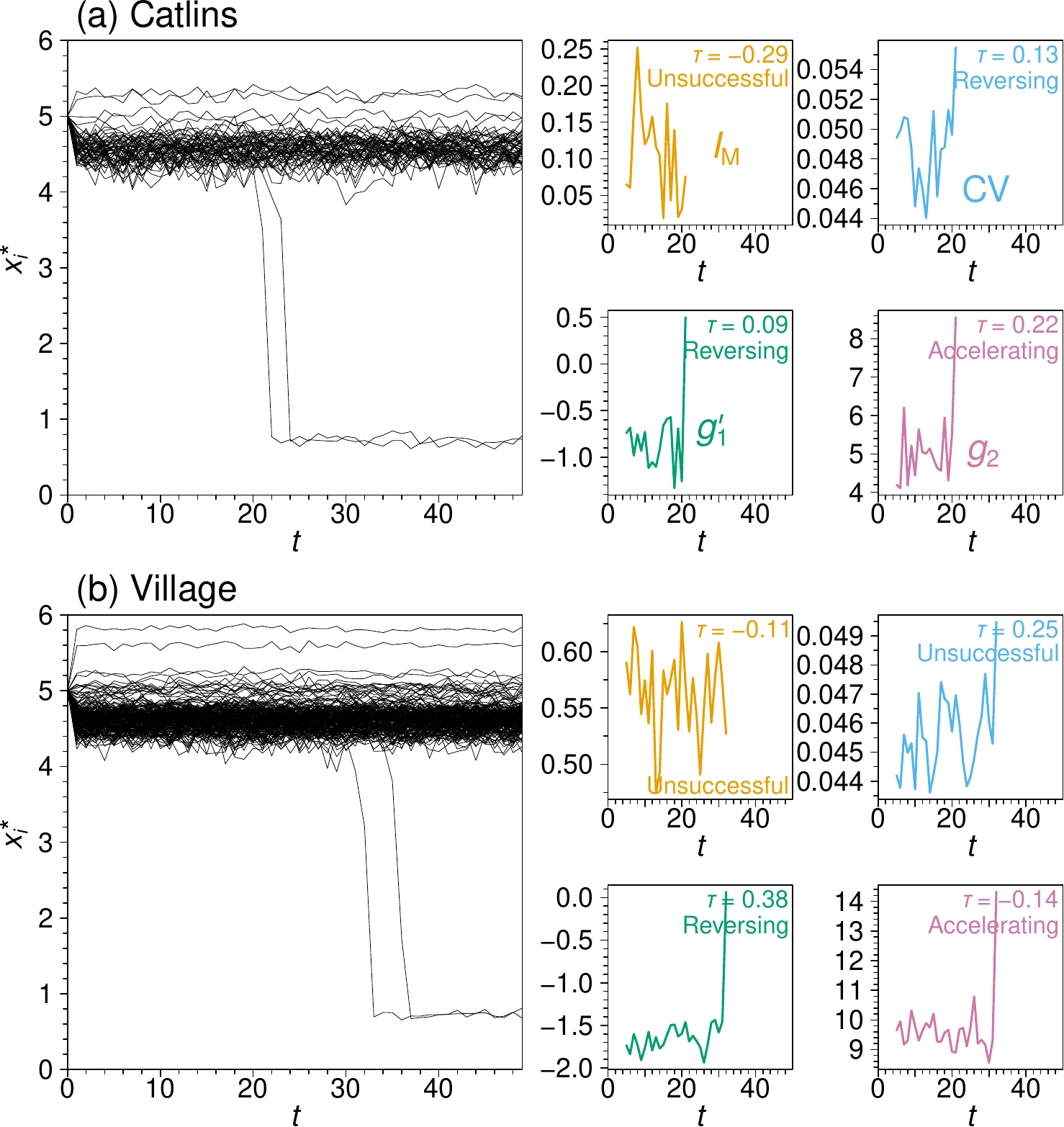}
  \caption{Node states and EWSs as a function of simulation time when dynamical noise induces a regime shift. We used the coupled double-well dynamics on two networks and do not change the parameter values during the course of the simulation. We show one $x_i(t)$ or EWS value per time unit. (a) Catlins network. (b) Village network.}
  \label{fig:example-stochastic}
\end{figure}

We find that the $\tau$ values for each EWS and each network are small in magnitude and unpredictable in sign, ranging from $-0.29$ to $0.38$ (see Fig.~\ref{fig:example-stochastic}). The time courses of EWS look highly noisy. Our classification scheme does not appear reliable in this scenario although the three out of the four EWSs are categorized as successful.

The second scenario is when a sudden large change in a control parameter forces a transition of some nodes. We use the same two networks and set $D=0.05$ and $\sigma = 0.01$. Initially we set $x_{i, 0} = 5$ and $u=-2$. At $t=50$, we set $u=-4$, which causes a tipping cascade in both networks without involving critical slowing down. We run the simulation for an additional 50 time units to observe further changes in the system state. As we have done in the first scenario, we take one sample from $x_i(t)$, $\forall i$ with which to compute EWS.

\begin{figure}[b]
  \includegraphics[width = 0.7\textwidth]{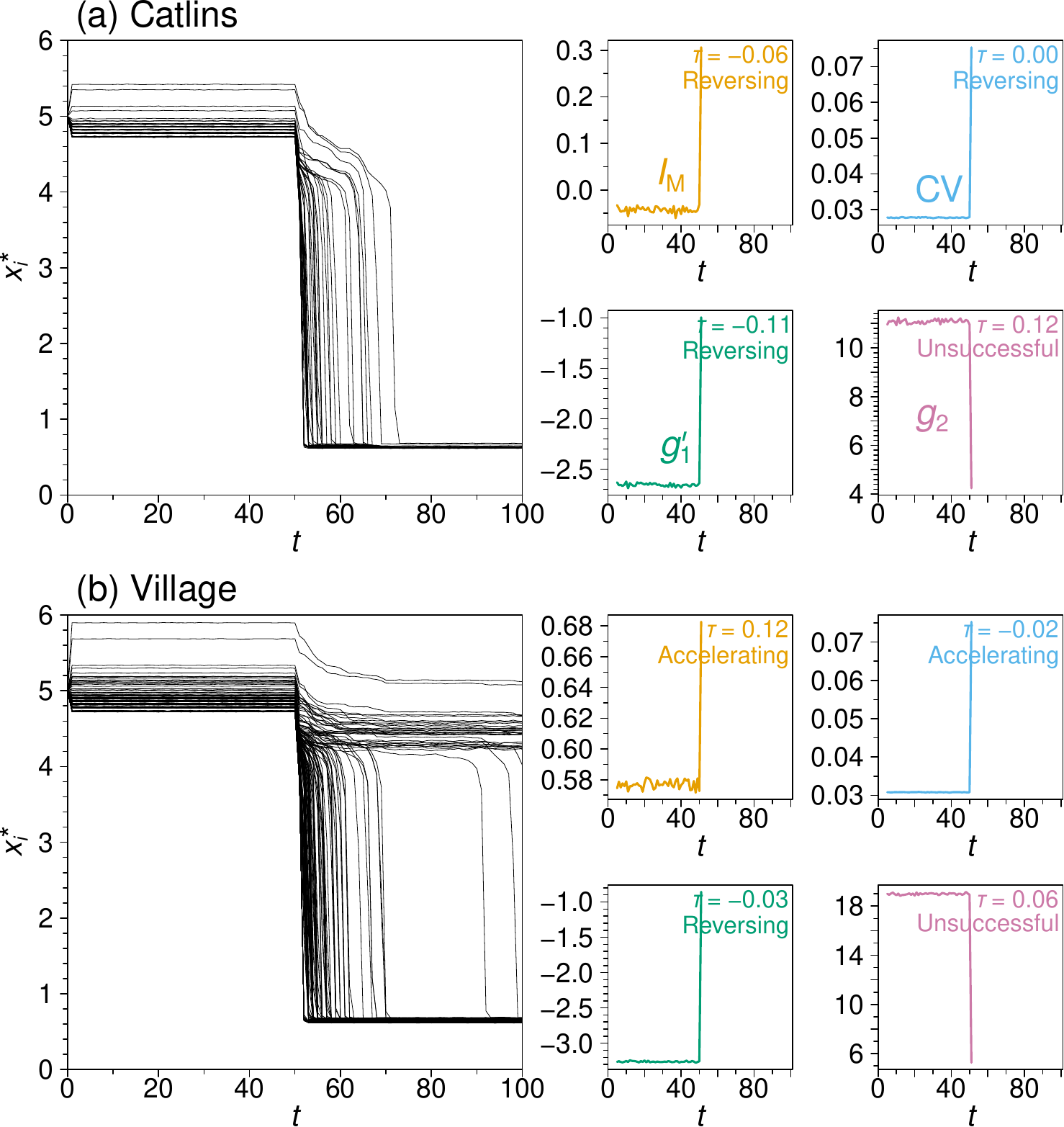}
  \caption{Node states and EWSs as a function of time when a sudden change in the stress parameter causes a regime shift. We use the coupled double-well dynamics, initially setting $u=-2$. We set $u=-4$ at $t=50$. (a) Catlins network. (b) Village network.}
  \label{fig:example-impulse}
\end{figure}

In this scenario, we find that all four EWSs fluctuate noisily around an initial value, then change suddenly and dramatically as some nodes happen to catch the beginning of the tipping cascade (see Fig.~\ref{fig:example-impulse}). The $\tau$ values are again small in magnitude and have an unpredictable sign, ranging from $-0.11$ to $0.12$. The EWSs remain flat until when the tipping cascade occurs. Our classification algorithm again appears unreliable.

We conclude that, when there is no critical slowing down, at least in these two scenarios, none of the four EWSs provides reliable indication of the impending transition prior to it.

%\clearpage

\end{document}